\documentclass[%
 reprint,
 showpacs,
 aps,
 superscriptaddress,
 pra,]{revtex4-2}

\usepackage{bm}% bold math
\usepackage{amsmath}
\usepackage{amssymb}
\usepackage{amsfonts}
\usepackage{braket}
\usepackage[dvipdfmx]{graphicx}
\usepackage{color}
\usepackage{here}

\begin{document}

\title{Multimode Quantum Correlations in Supercontinuum Pulses}

\author{Aruto Hosaka}
\affiliation{%
Mitsubishi Electric Corporation, Information Technology R\&D Center, Kanagawa 247-8501, Japan
}%
\affiliation{%
Department of Electronics and Electrical Engineering, 
Keio University, 3-14-1 Hiyoshi, Kohoku-ku, Yokohama, 223-8522, Japan
}%
\author{Shintaro Niimura}
\affiliation{%
Department of Electronics and Electrical Engineering, 
Keio University, 3-14-1 Hiyoshi, Kohoku-ku, Yokohama, 223-8522, Japan
}%
\author{Masaya Tomita}
\affiliation{%
Department of Electronics and Electrical Engineering, 
Keio University, 3-14-1 Hiyoshi, Kohoku-ku, Yokohama, 223-8522, Japan
}%
\author{Akihito Omi}
\affiliation{%
Department of Electronics and Electrical Engineering, 
Keio University, 3-14-1 Hiyoshi, Kohoku-ku, Yokohama, 223-8522, Japan
}%
\author{Masahiro Takeoka}
\affiliation{%
Department of Electronics and Electrical Engineering, 
Keio University, 3-14-1 Hiyoshi, Kohoku-ku, Yokohama, 223-8522, Japan
}%
\author{Fumihiko Kannari}
\affiliation{%
Department of Electronics and Electrical Engineering, 
Keio University, 3-14-1 Hiyoshi, Kohoku-ku, Yokohama, 223-8522, Japan
}

\date{\today}

\begin{abstract}
Suprecontinuum (SC) light contains complex spectral noise structure and its accurate characterization is important for fundamental understanding of its physics as well as for its applications. 
Several experimental and theoretical noise characterizations have been performed so far. However, none of them takes into account the quantum mechanical properties. 
Here, we demonstrate experimental characterisation of quantum noise and its spectral correlations formed in the SC light generated from a photonic crystal fiber.
Moreover, by applying an appropriate basis transformation to these correlations, we demonstrate that the SC noise amplitude can be squeezed below the shot-noise limit in some bases, even in the presence of excessively large nonlinearities.

%The spectral noise characteristics of supercontinuum (SC) light must be evaluated to determine the accuracy of SC light source experiments. 
%To determine the accuracy of SC light source experiments, its spectral noise characterization is a critical issue.  
%To obtain a full understanding of SC noise characteristics, evaluation of quantum mechanical properties is needed while previous studies are mostly based on classical physics (measurements). 
%Here, we demonstrate experimental characterise of quantum correlations formed in the SC spectrum, which allows us to evaluate the noise characteristics arising from quantum physics. Moreover, by applying an appropriate basis transformation to these correlations, we demonstrate that the SC noise amplitude can be squeezed below the shot-noise limit, even in the presence of excessively large nonlinearities.

\end{abstract}
\maketitle

\section{Introduction}

Supercontinuum (SC), also called `white light', is a valuable tool for studies requiring broadband characteristics \cite{PCF}, such as infrared spectroscopy \cite{MIR}, optical coherence tomography (OCT) \cite{OCT}, pulse compression \cite{Compress}, and carrier-envelope-phase stabilization for an optical frequency comb \cite{OFC}. Such light sources can be generated in two ways: by inserting femtosecond pulses into highly nonlinear media (e.g. photonic crystal fibers [PCFs] \cite{SC_PCF}) or by applying filament propagation of amplified high-peak femtosecond pulse lasers in hollow-core fibers \cite{GAS}, bulk crystals \cite{filamentation} or free space and silica glass plates \cite{silica}). Both methods require a thorough knowledge of the spectral broadening process caused by third-order nonlinear effects.

In addition to its broadness, another important factor of the SC is its {\it coherence}. Some applications require coherence in SC light, while the others require incoherent SC sources. Several studies \cite{Coherence1,Coherence2,Coherence3,Coherence4,Coherence5,Coherence6,Coherence7} have investigated the influence of noise and nonlinear processes on SC coherence. Corwin {\it et al}. \cite{Coherence4} experimentally and theoretically investigated broadband noise generation in the SC and suggested that SC noise stems from the nonlinear amplification of quantum noise. Using experimental approaches, N\"{a}rhi {\it et al}. \cite{Coherence5} spectrally and temporally characterised the second-order coherence of the SC and demonstrated a trade-off between second-order coherence and spectral broadening; moreover, their results were supported by numerical simulations. This result indicates that noise increases (coherence decreases) during the nonlinear spectral broadening process.
The above work, however, take into account only electromagnetics (i.e. classical theory) and a quantum mechanical understanding of second-order coherence in SC pulses is still lacking.  

By contrast, in quantum optics literature, fiber nonlinearity has been shown to {\it reduce} quantum noise in some physical observables. Photon-number squeezing (i.e. the suppression of photon-number fluctuations below the shot noise) has been observed for spectrally filtered soliton pulses \cite{soliton_filter} and SC pulses \cite{Hirosawa} propagating through nonlinear fibers. This behaviour is caused by Kerr nonlinearity in the fiber, which induces nontrivial intra-pulse quantum correlations among the spectrum of ultrafast optical pulses. Sp\"{a}lter {\it et al.} \cite{Soliton} experimentally measured the spectral quantum correlations generated via nonlinear fiber propagation, and this report was followed by further theoretical works \cite{ModeStructure}. Experimental modal analysis has been performed for a multimode-squeezed vacuum generated by parametric down-conversion with homodyne detection \cite{WDM} and photon number resolving detector \cite{Wakui-san}. These studies suggest that the quantum noise structure of an SC pulse consists of more than simply additional induced noise \cite{Coherence4}.

Recently, Ng {\it et al.} proposed a Gaussian split-step Fourier method to simulate quantum correlations formed within a supercontinuum \cite{GSSF}. Their simulations, especially in the context of supercontinuum generation based on saturated second-harmonic generation, revealed intricately patterned spectral correlations spanning an octave in frequency.

\begin{figure}[H]
  \begin{center}
    \includegraphics[clip, width=8.4cm]{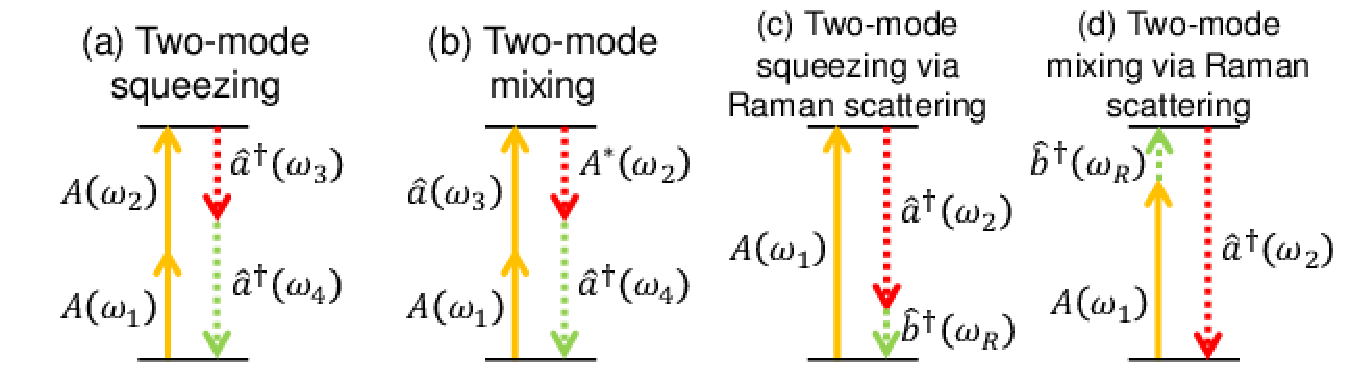}    
    \caption{Quantum phenomena of nonlinear propagation as a pulse travels through a fiber. $\hat{A}(\omega)$, $\hat{a}(\omega)$, $\hat{b}(\omega)$ represent the classical electrical field amplitude, the electrical field perturbation term and the phonon amplitude in the fiber, respectively. Two-mode squeezing induces an entanglement between two modes, and squeezed light is created when the two modes are degenerate. Two-mode mixing is analogous to the mixing operation in a beam splitter: two different frequency modes are mixed, and a phase shift is induced when they are degenerate.}
    \label{fig1}
  \end{center}
\end{figure}

\begin{figure*}[t]
  \begin{center}
    \begin{tabular}{c}

      % 1
      \begin{minipage}{0.95\hsize}
        \begin{center}
          \includegraphics[clip, width=15cm]{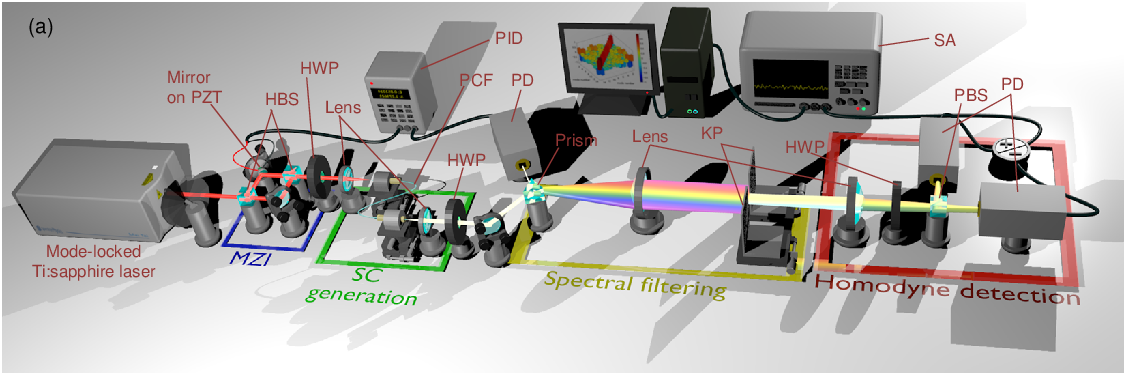}
        \end{center}
      \end{minipage}
	\\
      % 2
      \begin{minipage}{0.95\hsize}
        \begin{center}
          \includegraphics[clip, width=15cm]{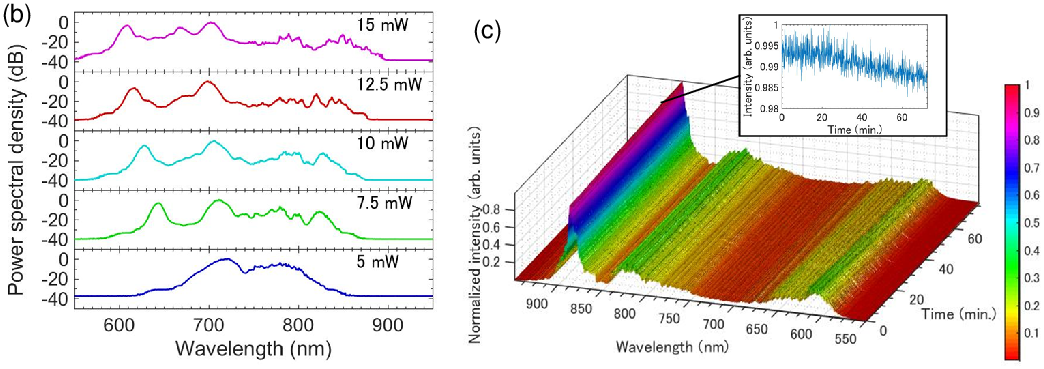}
        \end{center}
      \end{minipage}
    \end{tabular}
 
    \caption{Experimental setup for measuring quantum correlations in SC light. The Mach-Zehnder interferometer (MZI) stabilises the input pulse power and the spectral broadening process in the PCF. A knife edge filters the spectrum of the spectrally broadened pulse, and a homodyne measurement gives the photon-number noise and shot noise. PZT, piezo electric transducer; HBS, half-beam splitter; HWP, half-wave plate; PD, photodiode; KP, knife-edge pair; PBS, polarization beam splitter; SA, spectrum analyser. (b) Spectral power density for each input pulse power. We obtained results in 2.5-mW intervals between 5 mW and 15 mW. (c) Evaluation of the spectral stability for a 15-mW input pulse. This figure shows the change observed in the spectrum over a measurement of 76 min. The inset shows the time evolution of the spectral peak power, where the variation is within 2\%.}
    \label{fig2}
  \end{center}
\end{figure*}

In this paper, we experimentally characterise second-order photon-number quantum correlations in SC light generated via a nonlinear PCF. We measure the full information of second-order photon-number correlations among 19 binned spectral modes of the pulse. This information and a quantum mechanical modal analysis \cite{GQI} allow us to elucidate the more complex quantum structure of SC pulses. Specifically, by choosing a proper mode expansion in the time-frequency domain, we demonstrate that the SC pulse contains photon-number-squeezed modes, i.e. the noise intensity is reduced below the shot-noise limit, whereas the noise of the total pulse is increased, as previously reported in the SC literature. The multimode quantum noise structure revealed herein may open up new applications for SC light and may enable enhanced performance for current SC light applications.

\section{Nonlinear interactions of optical pulses in fiber}

The primary nonlinear processes that result in SC light generation in a fiber are four-wave mixing due to Kerr nonlinearities and Raman scattering. Let $\hat{A}(\omega)$ be the annihilation operator for frequency $\omega$. Because we are considering the quantum fluctuation of the strong (macroscopic) mean field in an optical pulse,  it is reasonable to approximate the pulse as $\hat{A}(\omega) \approx A(\omega) + \hat{a}(\omega)$, where $A(\omega) = \langle \hat{A}(\omega) \rangle$ is the expectation and $\hat{a}(\omega)$ is the perturbative annihilation operator. Then, the quantum noise evolution is dominated by the four nonlinear interactions illustrated in Fig.~1 (a)-(d): (a) two-mode-squeezing-like and (b) mode-mixing interactions between optical modes via Kerr nonlinearities and (c) two-mode-squeezing-like and (d) mode-mixing interactions between optical and vibrational modes via Raman scattering. The last two Raman scattering processes increase noise levels due to coupling with environmental phonons. The first process, however, induces squeezing among two optical modes, while the optical mode mixing in the second process does not change the noise level. Intuitively, by properly resolving the mixed mode, one can identify the squeezed modes involved in the SC pulse.
 
 \begin{figure*}[t]
  \begin{center}
    \begin{tabular}{c}

      % 1
      \begin{minipage}{0.45\hsize}
        \begin{center}
         \includegraphics[clip, width=8cm]{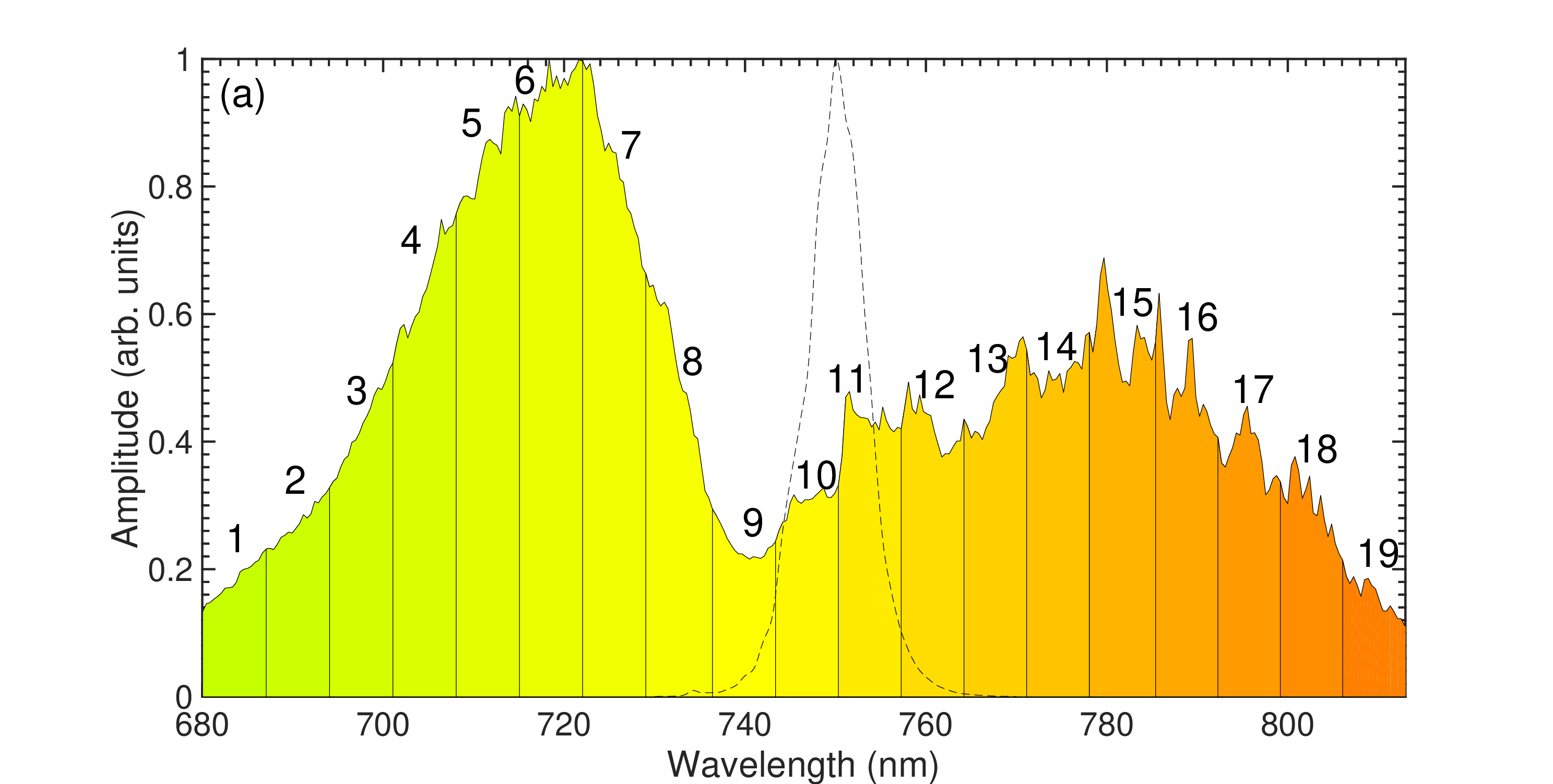}
        \end{center}
      \end{minipage}

      % 2
      \begin{minipage}{0.45\hsize}
        \begin{center}
          \includegraphics[clip, width=8cm]{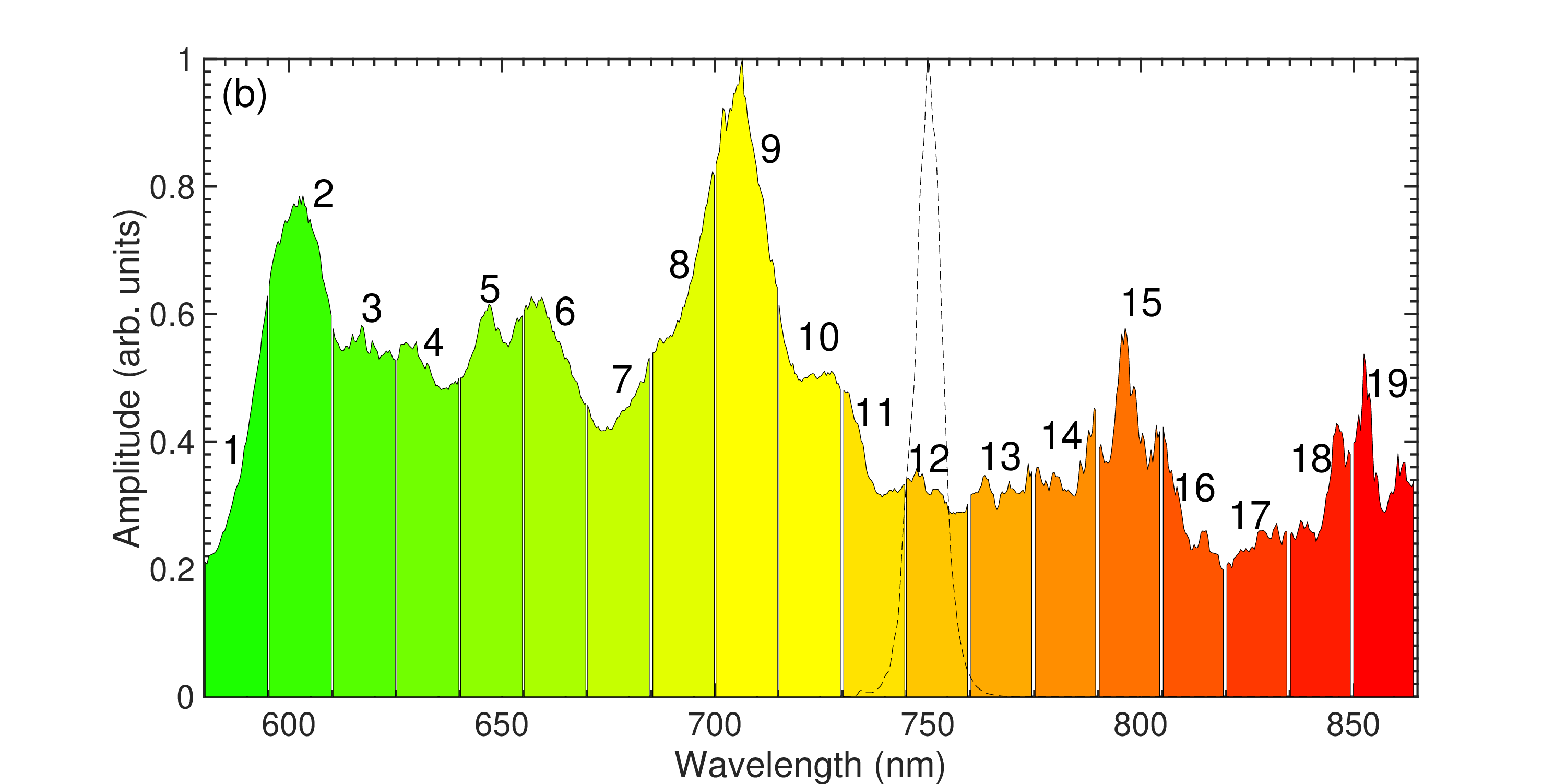}
        \end{center}
      \end{minipage}\\

      % 3
      \begin{minipage}{0.45\hsize}
        \begin{center}
          \includegraphics[clip, width=8cm]{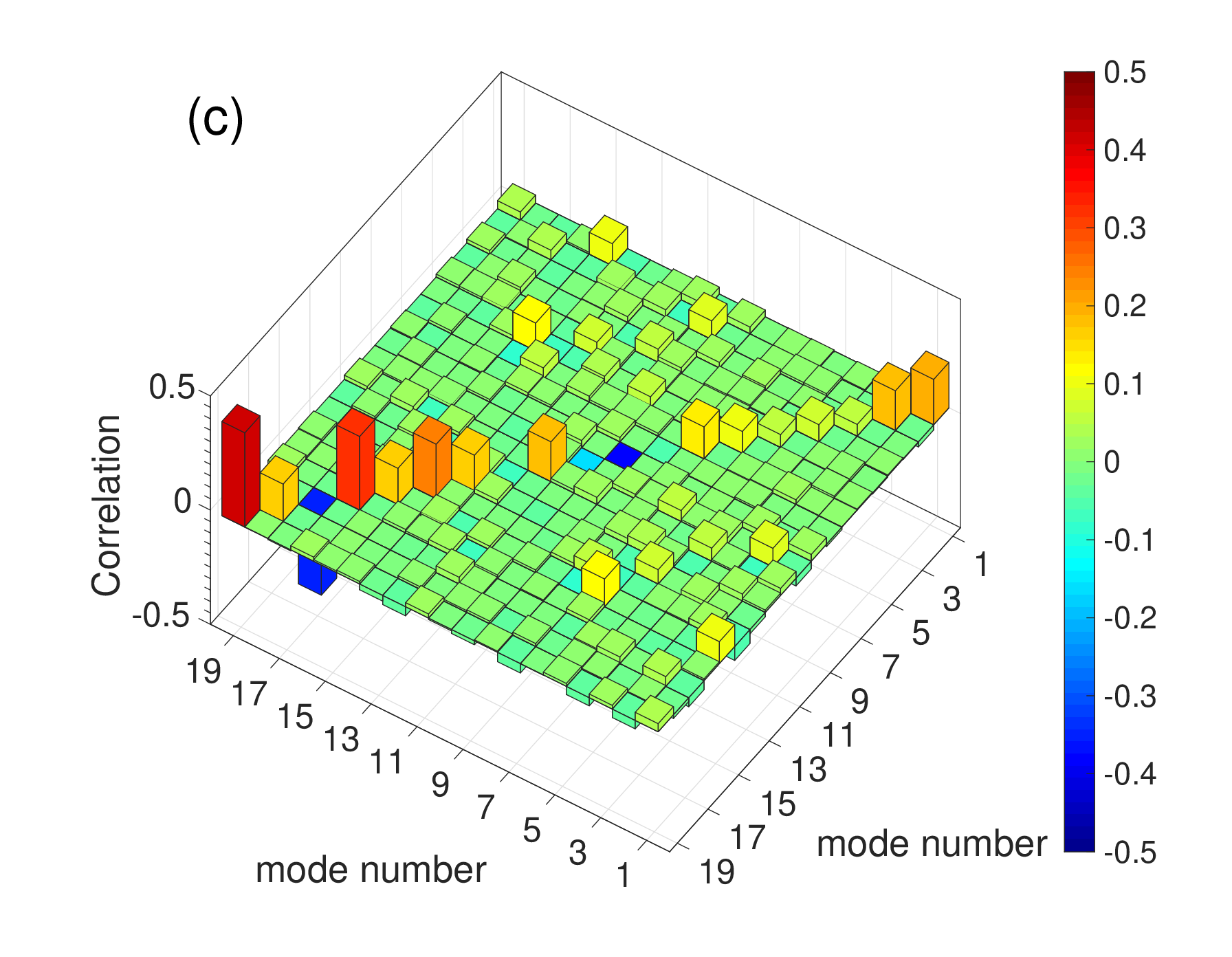}
        \end{center}
      \end{minipage}

      % 4
      \begin{minipage}{0.45\hsize}
        \begin{center}
          \includegraphics[clip, width=8cm]{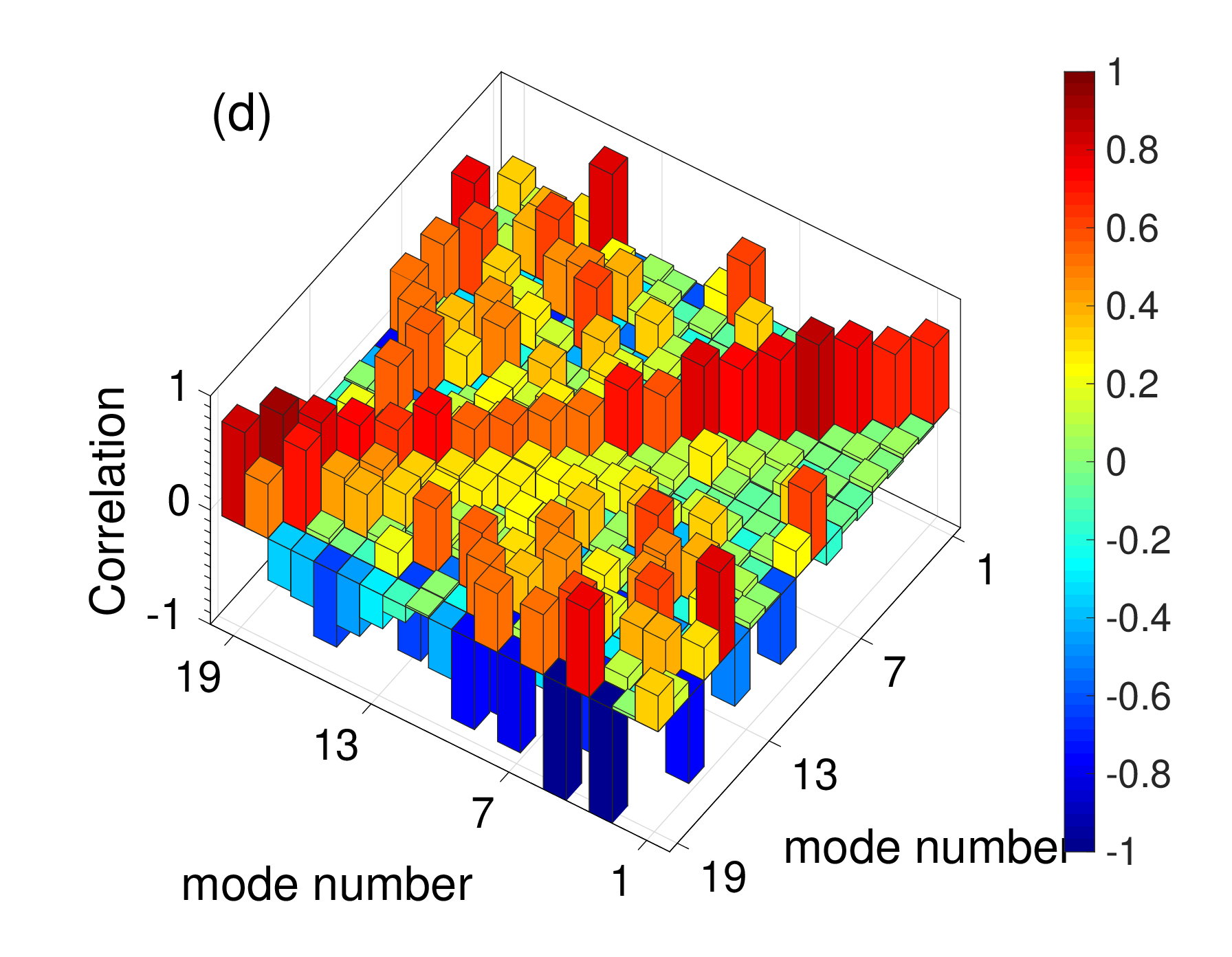}
        \end{center}
      \end{minipage}
    \end{tabular}
    
    \caption{Spectral amplitude of the output pulse for an incident pulse power of (a) 5 mW and (b) 15 mW. The dotted lines show the spectra of the input pulse. We divided these pulses into 19 spectral components. (c) and (d) present the measured covariance matrices among the spectral components shown in (a) and (b), respectively.}
    \label{fig3}
  \end{center}
\end{figure*}

\begin{figure*}[t]
  \begin{center}
    \begin{tabular}{c}

      % 1
      \begin{minipage}{0.5\hsize}
        \begin{center}
          \includegraphics[clip, width=9cm]{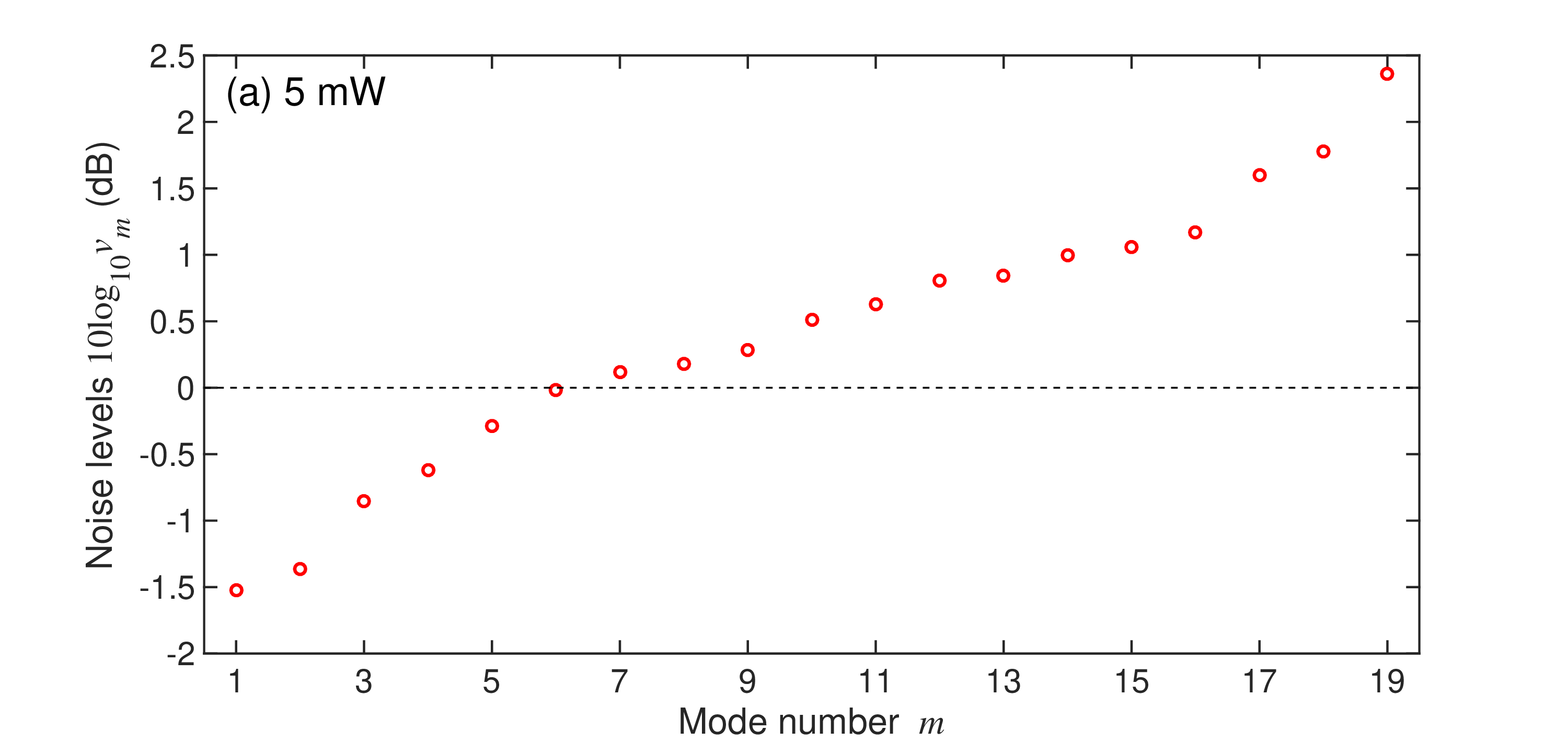}
        \end{center}
      \end{minipage}

      % 2
      \begin{minipage}{0.5\hsize}
        \begin{center}
          \includegraphics[clip, width=9cm]{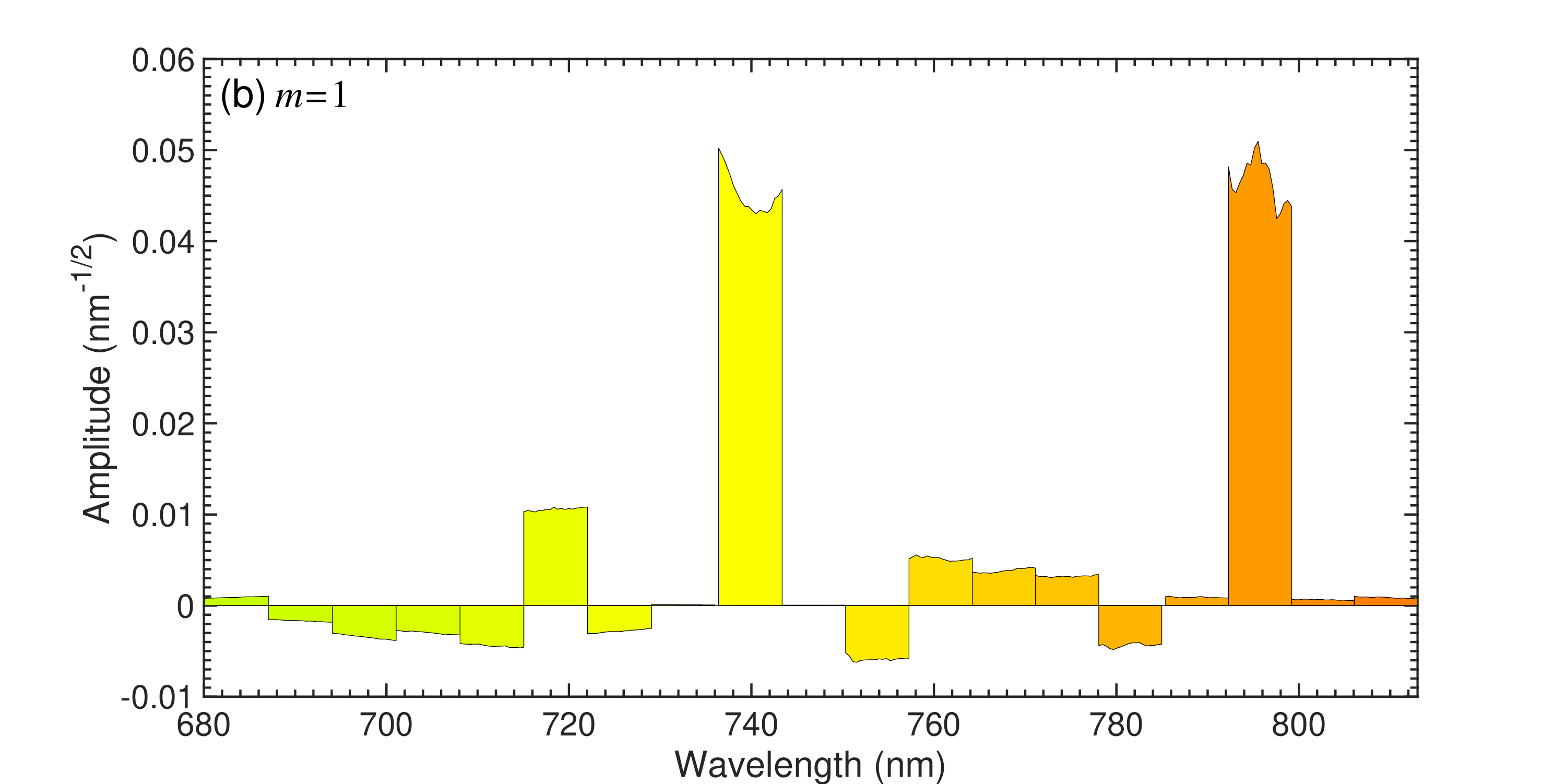}
        \end{center}
      \end{minipage}\\

      % 3
      \begin{minipage}{0.5\hsize}
        \begin{center}
          \includegraphics[clip, width=9cm]{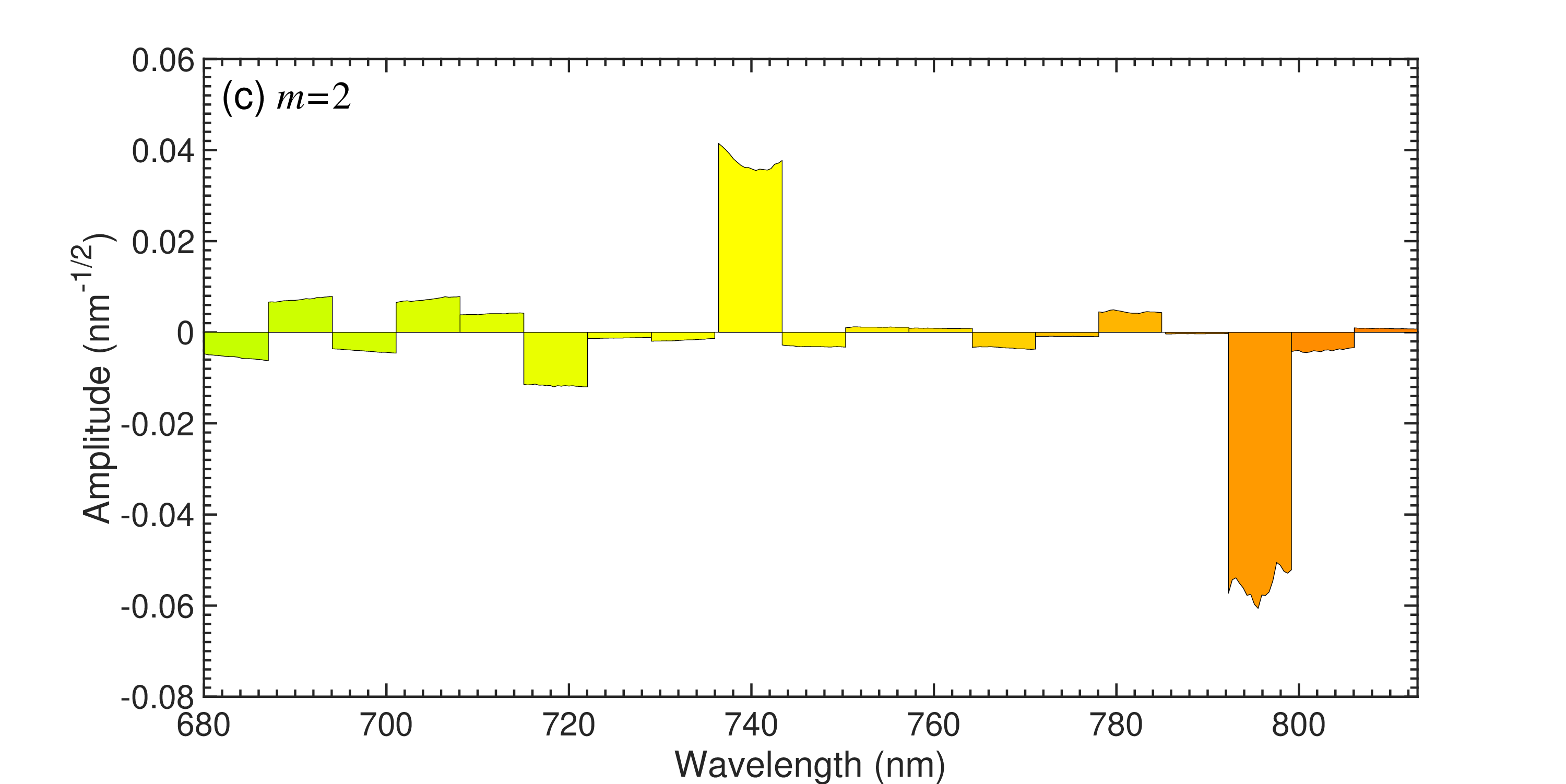}
        \end{center}
      \end{minipage}

      % 4
      \begin{minipage}{0.5\hsize}
        \begin{center}
          \includegraphics[clip, width=9cm]{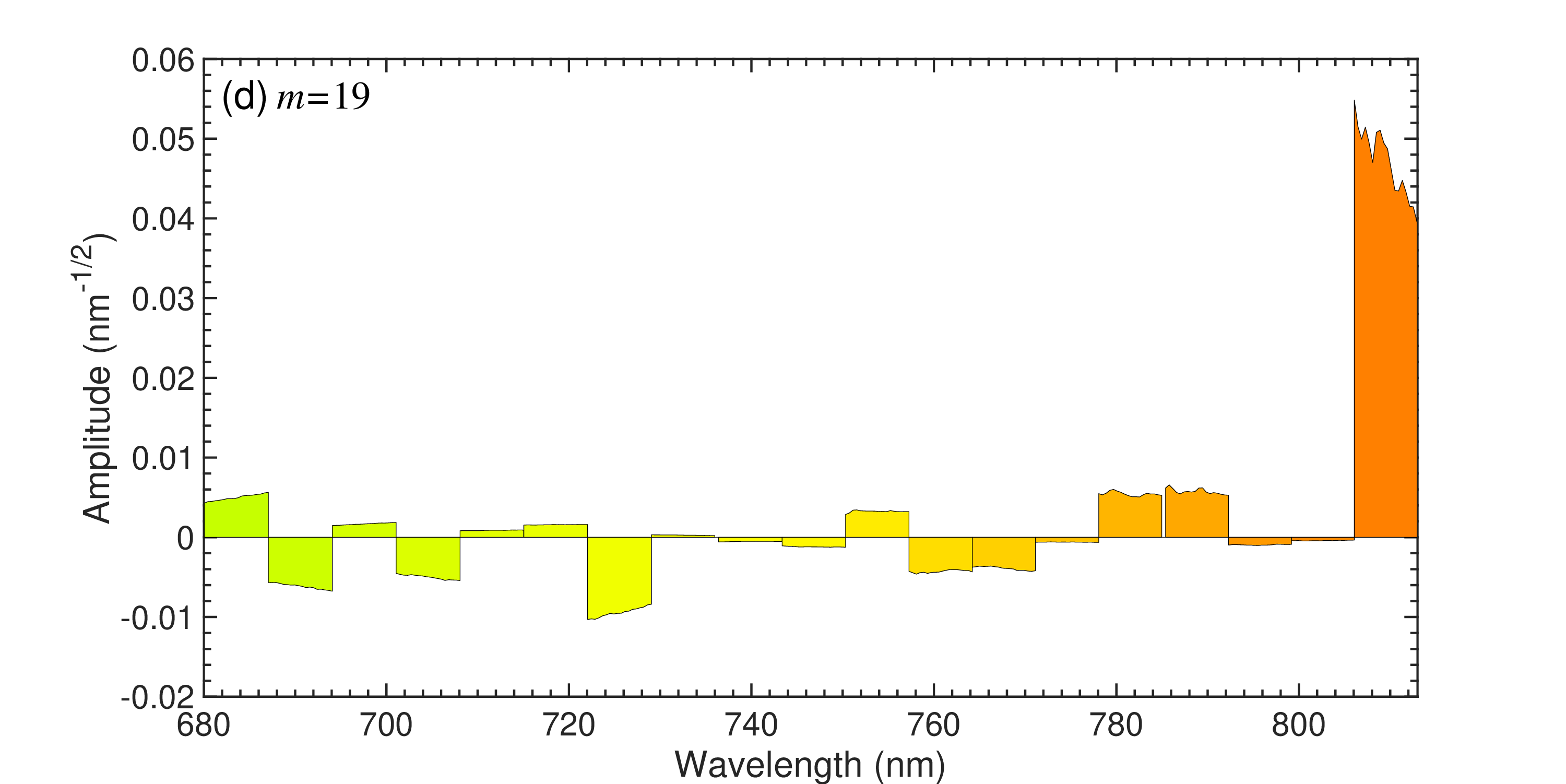}
        \end{center}
      \end{minipage}
    \end{tabular}
    
    \caption{Diagonalized covariance matrix for the SC light pumped by 5-mW pulses. (a) shows the squeezing level (noise level normalized by the shot noise) obtained for each eigenmode ($10\mathrm{log}_{10}{\it{v_m}}$). (b-d) display the spectral amplitude of the eigenmode for the $m=1,2,19$ basis states.}
    \label{fig4}
  \end{center}
\end{figure*}

\begin{figure*}[t]
  \begin{center}
    \begin{tabular}{c}

      % 1
      \begin{minipage}{0.5\hsize}
        \begin{center}
          \includegraphics[clip, width=9cm]{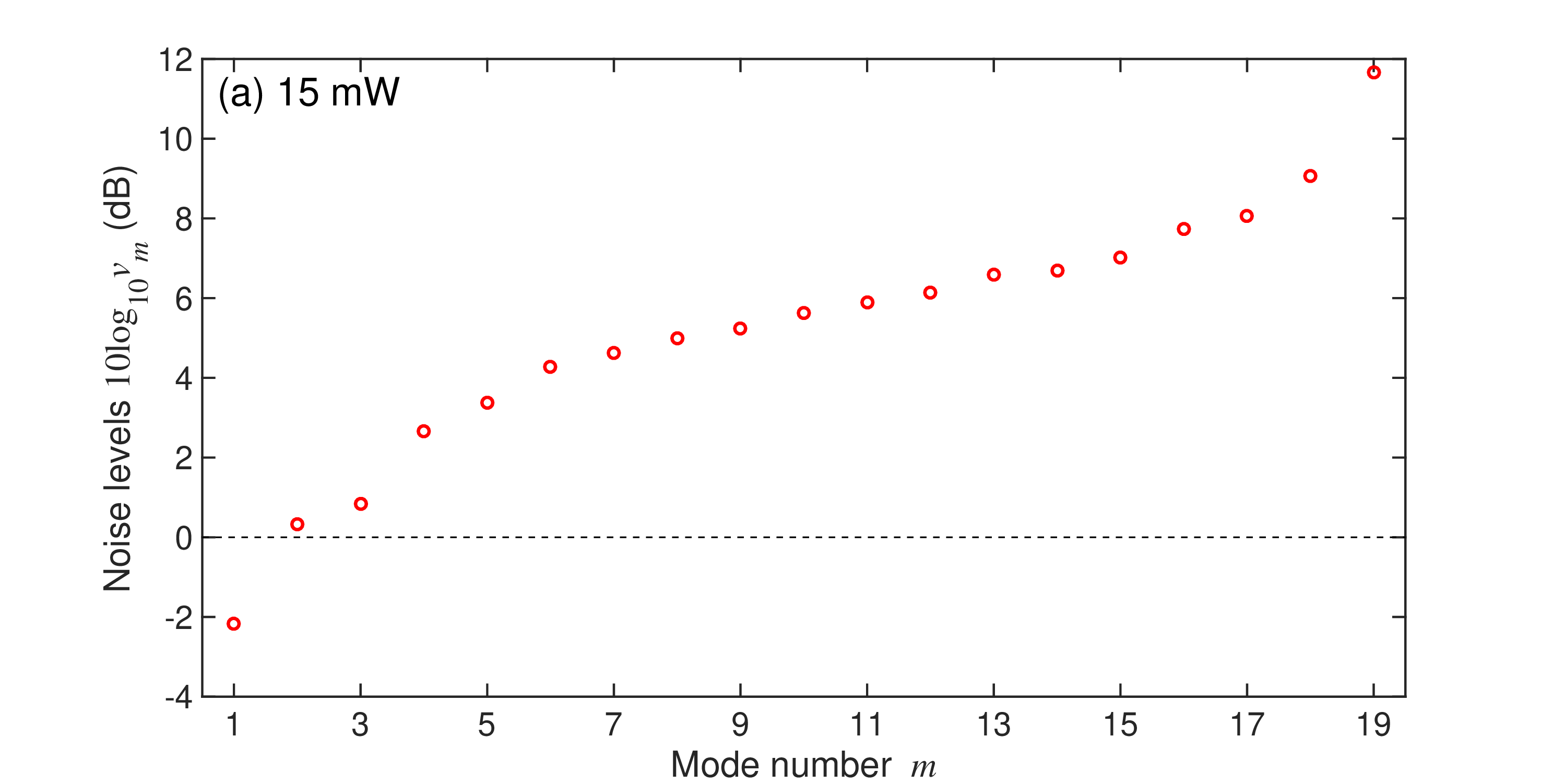}
        \end{center}
      \end{minipage}

      % 2
      \begin{minipage}{0.5\hsize}
        \begin{center}
          \includegraphics[clip, width=9cm]{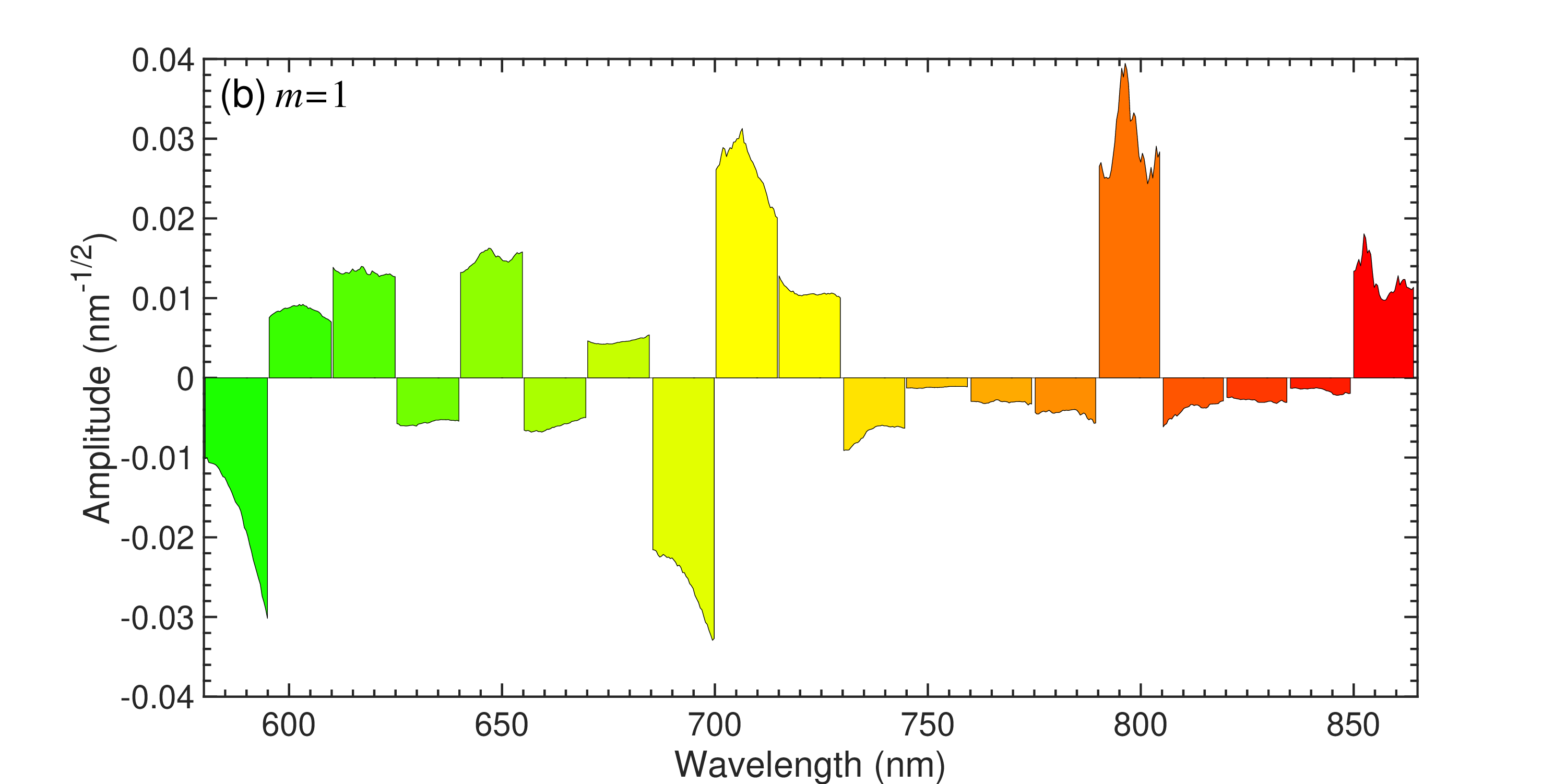}
        \end{center}
      \end{minipage}\\

      % 3
      \begin{minipage}{0.5\hsize}
        \begin{center}
          \includegraphics[clip, width=9cm]{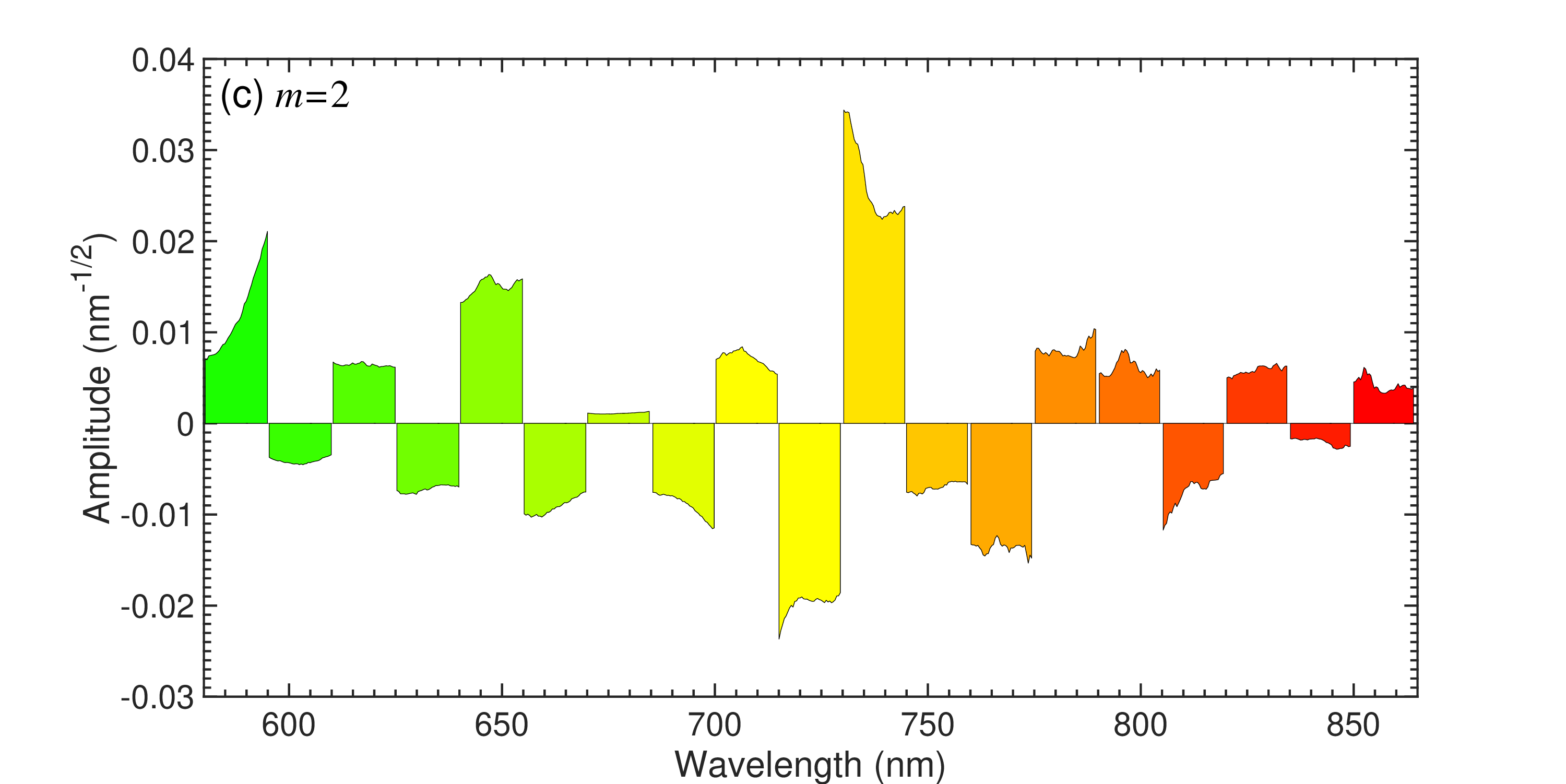}
        \end{center}
      \end{minipage}

      % 4
      \begin{minipage}{0.5\hsize}
        \begin{center}
          \includegraphics[clip, width=9cm]{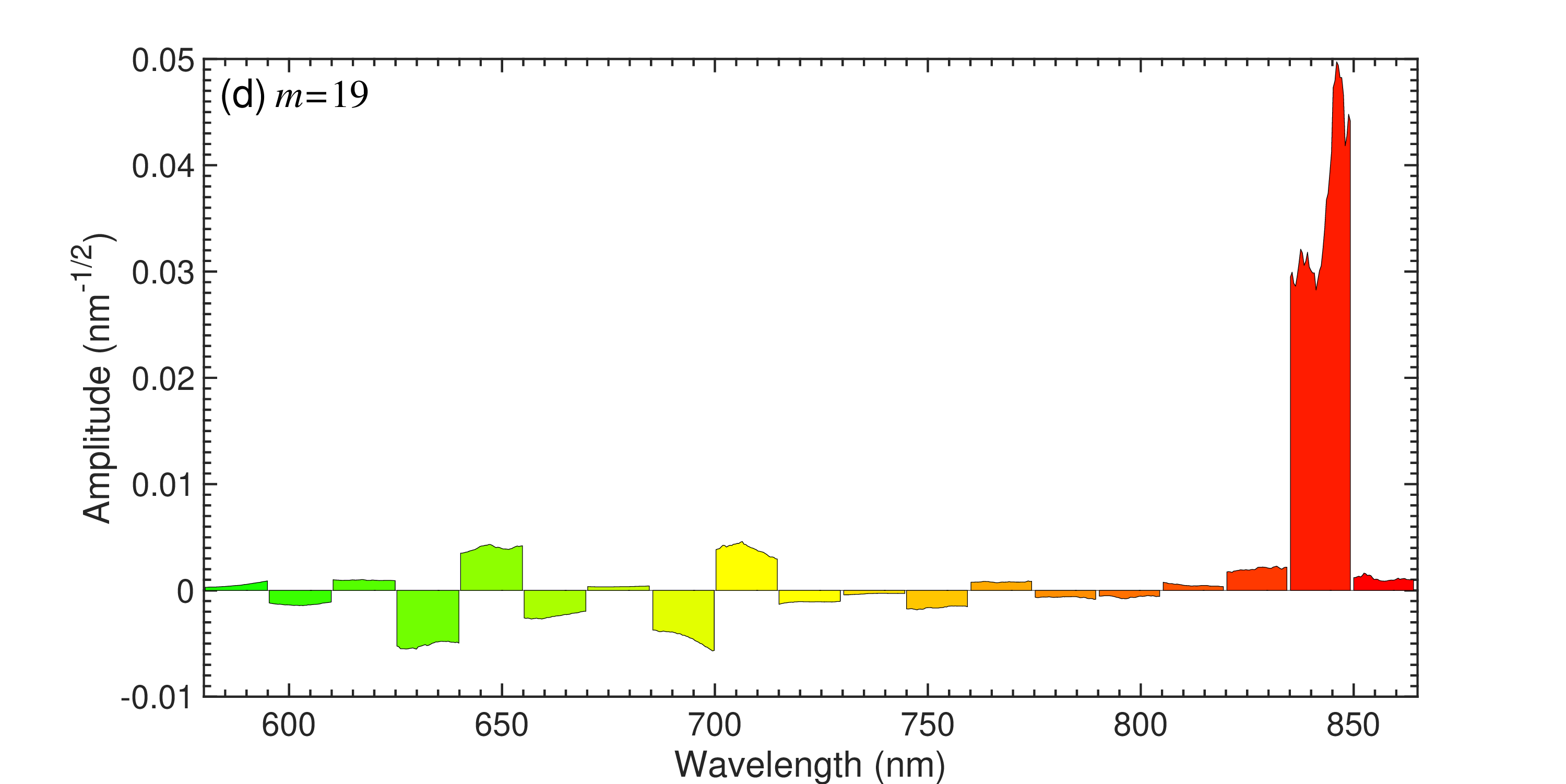}
        \end{center}
      \end{minipage}
    \end{tabular}
    
    \caption{Diagonalized covariance matrix for the SC light pumped by 15-mW pulses. (a) shows the squeezing level (noise level normalized by the shot noise) obtained for each eigenmode ($10\mathrm{log}_{10}{\it{v_m}}$). (b-d) display the spectral amplitude of the eigenmode for the $m=1,2,19$ basis states.}
    \label{fig5}
  \end{center}
\end{figure*}

\section{Experimental measurement of photon-number correlations in SC pulses}
Our experiments were conducted using the setup shown in Fig.~2(a). A mode-locked Ti:sapphire laser (SpectraPhysics, MaiTai-HP) with a repetition frequency of 80 MHz, a centre frequency of 750 nm and a pulse width of 90 fs was used. To generate SC light, the pulse was inserted into a 20-cm polarized, highly nonlinear PCF (NKT Photonics, NL-PM-750). The core diameter of the fiber was 1.8 $\mathrm{\mu m}$, and the zero-dispersion wavelength corresponded to the centre frequency of the laser pulse ($\beta_2=0\,\mathrm{ps^2/km}$, $\beta_3=0.0638\,\mathrm{ps^3/km}$, $\beta_4=-2.895\times10^{-5}\,\mathrm{ps^4/km}$). The input pulse polarization was aligned with the fiber axis by using a half-wave plate, and the output pulse from the fiber was aligned with the horizontal polarization by using a broadband wave plate. The incident power dependency of the SC spectra is shown in Fig.~2(b). The SC light generated from the fiber was collimated with a lens and separated into wavelengths with a Brewster prism modified for a 750-nm wavelength. The SC spectrum and nonlinear effects in the fiber differ according to the input pulse power; thus, a system capable of providing a stable input pulse energy was introduced. This system used SC light reflected from the prism, and the pulse power was measured by a photodiode. The power signal was sent to a PID controller, which then sent the PID signal to a piezo transducer. In this manner, the phase interference of the Mach-Zehnder interferometer was controlled, creating a stable input power and, hence, a stable SC spectrum (Fig.~2(c)).

Each spectral component of the SC light was divided by a prism and then focused in the Fourier plane by a lens. A movable knife-edge pair was placed as a spectral filter in this Fourier plane, and a photon-number noise measurement was performed for the transmitted spectral components. The pulse was divided into two equivalent beams using a half-wave plate and a polarized beam splitter. The two beams were measured by photodiodes, and the signals were aggregated with a 180-degree hybrid junction. Photon-number signals were obtained from the sum of the two signals, and shot-noise signals were obtained from the difference between the two signals.

Frequency-dependent losses may cause unwanted changes in the photon-number correlation structures. To avoid these changes, we used achromatic lenses to resolve the chromatic aberration to the greatest extent possible. We also applied prisms, instead of diffraction gratings, for wavelength division to lessen any wavelength-dependent losses. Furthermore, for the pulse propagating after the fiber, we used silver mirrors and a wavelength plate that covers a broadband spectrum.

The generated SC pulse was divided into 19 spectral bins, and we measured the covariance matrix of the photon-number fluctuations, $\Braket{\hat{n}_m , \hat{n}_{m'} } \equiv \Braket{\hat{n}_m \hat{n}_{m'}}-\Braket{\hat{n}_m}\Braket{\hat{n}_{m'}}$, where $\hat{n}_m$ is the photon number operator for the $m$th bin (spectral mode). For this step, we first measured the photon-number autocorrelation of the spectrally filtered pulse. The knife edges can filter out the low- and high-frequency modes of the pulse and transmit from the $k$th to $(k+l)$th modes with the spectral window of $T=\{ k,k+1,\dots,k+l \}\,(1 \le k, l \le 19, \, k+l\leq19)$. 
The photon number fluctuation of the filtered pulse is then given by 
$\langle \left( \sum_{m \in T} \hat{n}_m \right)^2 \rangle - \left( \sum_{m \in T} \langle \hat{n}_m \rangle \right)^2 = \sum_{m,m' \in T} \langle \hat{n}_m, \hat{n}_{m'} \rangle$. 
By changing the knife edge window, $k$ and $l$, the photon number fluctuation of 190 combinations of $T$ can be obtained, and the simultaneous equation can be solved to give a 19 x 19 matrix consisting of $\langle \hat{n}_m, \hat{n}_{m'} \rangle$. \cite{Soliton}.

The matrix is further normalised as follows. Let $\hat{A}_m$ be the annihilation operator for the $m$th mode, and let us take the perturbative expansion as $A_m + \hat{a}_m$. By neglecting the second-order perturbation, we have $\hat{n}_m \approx N_m + |A_m| \hat{x}_m$, where $N_m$ and $A_m$ are the mean photon number and the complex mean quadrature amplitude of the $m$th mode, respectively, and $\hat{x}_m$ is the corresponding perturbative quadrature amplitude operator. Then, we obtain $\Braket{\hat{n}_m , \hat{n}_{m'} } \approx |A_m| |A'_m| \Braket{ \hat{x}_m \hat{x}_{m'}} \equiv |A_m| |A_{m'}| C_{m,m'}$.

Next, we refer to the normalised matrix $C=[ C_{m,m'} ]_{m,m'}$ as the covariance matrix. Note that the phase of $\hat{x}_m$ is identical to that of the mean field $A_m$.the squeezing along this quadrature approximately corresponds to the intensity (photon-number) squeezing.

We experimentally obtained the amplitude $|A_m|$ by measuring the shot-noise level for each spectral component. The results of the homodyne measurement are given by $|A_m|^2$  because the vacuum noise is normalised as 1. By measuring 19 shot-noise levels, we obtained the spectral amplitude $|A_m|$ and the normalized covariance matrix $C$ from $\Braket{\hat{n}_m , \hat{n}_{m'} }$.

The experimental results are shown in Fig.~3. Figures 3(a) and (b) show the spectra observed when pulses with an average power of 5 mW or 15 mW were inserted into the fiber, respectively. The dotted lines present the input spectrum, and the output is shown as the colored region. Although a broadband spectrum was not obtained for 5 mW, we measured the covariance matrix for comparison. The output was divided into 19 spectral bins (marked by different colors), and we obtained covariance matrices of the photon-number correlations from knife-edge filtering measurements.

The observed covariance matrices are depicted in Figs.~3(c) and (d). The diagonal term (autocorrelation) corresponds to the spectrally resolved photon-number fluctuation: positive values indicate increased noise while negative values correspond to photon-number squeezing. More noise is observed for the higher input (15 mW), which is consistent with the broadband noise increase observed in SC light generation. However, the situation becomes more complex when we carefully consider the nondiagonal terms (mutual correlation).

As mentioned in the previous section, the fiber nonlinearity contributes to the amplification of spectral quantum noise via two processes: the Raman scattering and the cross Kerr effect.
%mutual quantum correlations between different spectral modes caused by the Kerr effect. 
The former process is intrinsically unavoidable; however, the latter effect can be prevented by properly {\it demixing} the mutual quantum correlation, i.e. the nondiagonal terms of the covariance matrix (see \cite{GQI}, for example, for the covariance matrix approach for Gaussian quantum information processing). In other words, one can suppress this noise by choosing a proper mode basis in the frequency-time domain.

%\clearpage

We perform a mode transformation by diagonalising the covariance matrix:

\begin{eqnarray}
  C_{m,m'}=U^{\rm{T}}VU. \label{eq2}
\end{eqnarray}

Here $U$ is a $19\times19$ unitary matrix, and $V=\mathrm{diag}\{v_1,\dots,v_{19}\}$ $(v_1 \leq v_2 \leq \dots \leq v_{19})$ is a diagonal matrix in which each eigenvalue corresponds to an element. The unitary matrix $U$ corresponds to the mode transformation. From the frequency-mode basis $\hat{\bf x} \equiv (\hat{x}_1,\dots,\hat{x}_{19})^{\mathrm{T}}$ 
 we define the new basis as

\begin{eqnarray}
  (\hat{x}_1',\dots,\hat{x}_{19}')^{\mathrm{T}}=U\hat{\bf x}. \label{eq3}
\end{eqnarray}

In this new basis, each mode is uncorrelated in terms of the photon-number variance, $\Braket{\hat{x}_m' \hat{x}_{m'}'}=\delta_{m,m'}v_m$. $\hat{x}_1'$ and $\hat{x}_{19}'$ represent the minimum and the maximum noise levels, respectively, among all modes in the new basis.

Figures 4(a) and 5(b) show the covariance matrices for the diagonalized basis for the 5-mW and 15-mW pumping, respectively, where Figs.~4(b-d) and 5(b-d) illustrates the spectral amplitude of the eigenmode for $m=1, 2, 19$. We observe photon-number squeezing in the first five modes for the case of low pump power (5 mW). This observation agrees with our theoretical results \cite{Multimode} on multimode squeezing in the zero-dispersion fiber propagation. Interestingly, even for high pump power (15 mW), one mode still shows photon-number squeezing. 
In classical optics-based analysis \cite{GSSF}, the excess spectral noise was observed for the SC light. 
Here, on the other hand, we successfully obtained noise-suppressed SC pulses by applying quantum mechanical modal analysis to the spectral noise correlation.
%For the conventional experiment and numerical analysis in classical optics, the SC exhibits excess spectral noise, as expected. However, we successfully obtained noise-suppressed SC pulses by applying modal analysis to the spectral noise correlation.

Note that spectral resolution by the 19-component division is not sufficient to observe the fine structure of of the full covariance matrix. Also, this approach acted as a low-pass filter for the covariance matrix. 
Thus, we predict that a higher squeezing level could be observed if we made finer spectral divisions.
%In this experiment, we divided the SC pulse into 19 components. This spectral resolution was insufficient to measure the fine structure of the full covariance matrix, and this approach acted as a low-pass filter for the covariance matrix. As a result, squeezing and anti-squeezing levels were degraded. Thus, we predict that a higher squeezing level can be obtained by measuring the fine structure of the covariance matrix. Conversely, we have shown that squeezed SC pulses can be obtained even with relatively poor spectral resolution and without a high-resolution measurement of the covariance matrix.

\section{Conclusion and future applications}

We have experimentally demonstrated a method for producing photon-number-squeezed states from complicated correlations by performing a modal analysis of the quantum correlation produced by SC light generation. To the best of our knowledge, this study is the first experimental observation of quantum correlations within an SC light pulse, which were theoretically suggested in \cite{Hirosawa}. 
While previous works \cite{Coherence1,Coherence2,Coherence3,Coherence4,Coherence5} have considered the SC noise only classically and in time or frequency bases, we have demonstrated that one can suppress the noise below the shot-noise limit by wisely  'demixing' the quantum correlation among the spectral components with proper time-frequency bases. 
%a carefully selected time-frequency basis can `demix' the quantum correlation and can suppress the noise in each mode, with some modes being below the shot-noise limit.

Our approach is beneficial for applications of SC light. For example, in conventional molecular spectroscopy, spectral components of SC pulses are generally treated separately, which renders excess noise at each frequency unavoidable. However, with our approach, by properly demixing the quantum correlations among spectral modes, one can suppress this noise and achieve sub-shot-noise sensitivity in SC light spectroscopy.

Future work regarding this method may focus on real-time quantum correlation analysis. Although our method has a simple setup with a minimal number of detectors, our approach for quantum correlation analysis was indirect and time-consuming. An array-type detector combined with post-selection mode transformation may allow real-time direct measurements of squeezed light \cite{GQC}. A similar approach for SC light should enable applications to achieve high-efficiency, high-speed optical measurements.

\begin{acknowledgements}
%\textit{Acknowledgements.}\textemdash
This research was supported by JST CREST Grant Number JPMJCR1772 and Grant-in-Aid for JSPS Fellows 16J03900.
\end{acknowledgements}

\appendix
\section{Homodyne detection}
The photon-number noise was measured using a photodiode (S38833, Hamamatsu). A broadband half-wave plate and a broadband polarization beam splitter were used to divide the pulse into two pulses, with each pulse entering a photodiode. The diameter of each photodiode is 1.5 mm, and the incident light is designed to have a beam width of less than 1 mm for all relevant frequencies. The measured photocurrents are sent to the 180-degree hybrid junction, which gives an output of the sum and difference of the signals, corresponding to the photon-number noise and vacuum shot-noise level, respectively. In the difference signal measurements, a high common mode rejection ratio (CMRR) cannot be simultaneously obtained for all frequencies, due to the chromatic characteristics of the polarization beam splitter. Thus, for the 19 patterns of shot-noise measurements, the half-wave plate was adjusted for each wavelength to ensure that the CMRR exceeded 20 dB.

After the 180-degree hybrid junction, the measured photon-noise or shot-noise signal passes through a bandpass filter (19.2-23.6 MHz) and is then amplified by a low-noise amplifier before being sent to a radio-frequency spectrum analyser (RFSA). The RFSA was set to perform measurements for a center frequency of 20 MHz, with a resolution bandwidth of 10 kHz and a video bandwidth of 1 kHz, and the average of a 1-s measurement was used as the noise level.

\section{Spectral filtering}
The wavelength was filtered by a 4-$f$ optical system. A Brewster prism from SF14 divided incident light with a 3-mm beam diameter into its spatial wavelength modes, and a 300-mm-focal-length lens was applied to create a Fourier plane. At the Fourier plane, the beam diameter was roughly 100 $\mathrm{\mu m}$ for each wavelength, and the system was designed to maintain a wavelength resolution below 1.2 nm for all wavelengths. The second lens in the 4-$f$ optical system has a focus length of 100 mm, creating a beam diameter of 1 mm at the photodiode.

\section{Covariance matrix}
Here, we provide data on the covariance matrix obtained in our experiments. The electrical noise introduces errors in the homodyne measurement results. The signal-to-noise ratio of our homodyne measurement system was 41 dB for a 100 $\mathrm{\mu W}$ shot-noise measurement. We considered the third digit as the effective digit in the photon-number noise measurement system and obtained the photon-number covariance matrix by solving the simultaneous equation.

The photon-number covariance matrix for a 5-mW pulse entering into a PCF is given below.

\begin{widetext}
\begin{align*}
  {\it{C}} = \tiny\left(
    \begin{array}{cccccccccc}
		1.2481&-0.022258&-0.0029839&0.004318&-0.0048727&0.0042037&0.0014184&-0.01054&0.029623&-0.0336\\
		-0.022258&1.2103&-0.021905&-0.0011658&0.011553&0.011098&0.0018055&-0.0013002&-0.029695&0.081569\\
		-0.0029839&-0.021905&1.0541&-0.018815&0.0024405&-0.0045643&0.020139&-0.0093263&0.024462&-0.0083288\\
		0.004318&-0.0011658&-0.018815&1.0813&-0.018617&0.0041863&0.011893&0.0087029&-0.0010979&0.025291\\
		-0.0048727&0.011553&0.0024405&-0.018617&1.06&-0.01721&0.0060853&0.0021074&0.0016679&0.008625\\
		0.0042037&0.011098&-0.0045643&0.0041863&-0.01721&1.1054&-0.017135&0.002417&-0.0040509&0.04799\\
		0.0014184&0.0018055&0.020139&0.011893&0.0060853&-0.017135&1.1628&-0.0095505&0.012605&0.011212\\
		-0.01054&-0.0013002&-0.0093263&0.0087029&0.0021074&0.002417&-0.0095505&0.94237&-0.0003657&0.019434\\
		0.029623&-0.029695&0.024462&-0.0010979&0.0016679&-0.0040509&0.012605&-0.0003657&0.72362&-0.0010198\\
		-0.0336&0.081569&-0.0083288&0.025291&0.008625&0.04799&0.011212&0.019434&-0.0010198&0.86441\\
		0.030553&-0.06321&0.0078075&0.057362&-0.02742&-0.0076014&0.011381&-0.0067318&0.030786&0.014449\\
		0.010277&0.030556&-0.013052&-0.054872&0.011182&0.038075&0.025568&-0.0058536&-0.013514&-0.0086106\\
		-0.018868&0.00052433&0.025277&-0.0095254&0.062633&-0.054032&-0.0071522&0.01135&-0.0074129&-0.04179\\
		-0.060892&0.025191&0.0054999&0.0025242&-0.032317&-0.071091&0.049982&0.016038&-0.017689&-0.010401\\
		0.09602&-0.041193&-0.0057176&-0.029277&-0.00027409&0.12163&-0.089916&-0.022402&0.011529&-0.007773\\
		0.0082153&-0.041375&-0.036672&-0.0055816&-0.017029&0.017645&-0.0045765&0.0048277&-0.0095821&0.014653\\
		-0.039507&0.042353&-0.0054583&0.029018&0.015708&-0.041467&-0.0015349&-0.0039517&-0.01545&0.0057668\\
		-0.016124&0.0033189&-0.018147&0.0049088&0.010307&-0.018279&0.013043&-0.036404&0.011423&0.0078501\\
		0.035652&-0.036715&0.016444&-0.035687&0.012817&-0.001113&-0.041131&0.0048778&-0.017117&-0.0061377
		\end{array}
    \right.\\\notag
	\tiny\left.
    \begin{array}{ccccccccc}
		0.030553&0.010277&-0.018868&-0.060892&0.09602&0.0082153&-0.039507&-0.016124&0.035652\\
		-0.06321&0.030556&0.00052433&0.025191&-0.041193&-0.041375&0.042353&0.0033189&-0.036715\\
		0.0078075&-0.013052&0.025277&0.0054999&-0.0057176&-0.036672&-0.0054583&-0.018147&0.016444\\
		0.057362&-0.054872&-0.0095254&0.0025242&-0.029277&-0.0055816&0.029018&0.0049088&-0.035687\\
		-0.02742&0.011182&0.062633&-0.032317&-0.00027409&-0.017029&0.015708&0.010307&0.012817\\
		-0.0076014&0.038075&-0.054032&-0.071091&0.12163&0.017645&-0.041467&-0.018279&-0.001113\\
		0.011381&0.025568&-0.0071522&0.049982&-0.089916&-0.0045765&-0.0015349&0.013043&-0.041131\\
		-0.0067318&-0.0058536&0.01135&0.016038&-0.022402&0.0048277&-0.0039517&-0.036404&0.0048778\\
		0.030786&-0.013514&-0.0074129&-0.017689&0.011529&-0.0095821&-0.01545&0.011423&-0.017117\\
		0.014449&-0.0086106&-0.04179&-0.010401&-0.007773&0.014653&0.0057668&0.0078501&-0.0061377\\
		1.2111&-0.023532&-0.029315&-0.0085247&0.016089&-0.068487&0.029452&-0.011159&0.018162\\
		-0.023532&0.93039&0.019523&-0.0020031&-0.050909&0.021435&-0.020944&0.011328&-0.039911\\
		-0.029315&0.019523&1.1981&0.011383&0.018461&0.0089306&-0.018118&-0.015371&-0.025214\\
		-0.0085247&-0.0020031&0.011383&1.313&-0.012341&0.014854&-0.0042785&-0.0067265&0.011191\\
		0.016089&-0.050909&0.018461&-0.012341&1.1875&-0.022749&0.0025607&8.565\times10^{-5}&0.0071284\\
		-0.068487&0.021435&0.0089306&0.014854&-0.022749&1.4756&-0.024503&-0.0011098&0.01811\\
		0.029452&-0.020944&-0.018118&-0.0042785&0.0025607&-0.024503&0.74195&-0.019167&-0.0044517\\
		-0.011159&0.011328&-0.015371&-0.0067265&8.565\times10^{-5}&-0.0011098&-0.019167&1.1955&0.0010146\\
		0.018162&-0.039911&-0.025214&0.011191&0.0071284&0.01811&-0.0044517&0.0010146&1.7031
	\end{array}\right)
\end{align*}

By performing an eigenvalue decomposition, we find

\begin{align*}
  {\it{V}} = \tiny\left(
    \begin{array}{cccccccccc}
		0.70382&0&0&0&0&0&0&0&0&0\\ 
		0&0.73046&0&0&0&0&0&0&0&0\\ 
		0&0&0.82163&0&0&0&0&0&0&0\\ 
		0&0&0&0.86729&0&0&0&0&0&0\\ 
		0&0&0&0&0.93585&0&0&0&0&0\\ 
		0&0&0&0&0&0.99563&0&0&0&0\\ 
		0&0&0&0&0&0&1.0275&0&0&0\\ 
		0&0&0&0&0&0&0&1.0424&0&0\\ 
		0&0&0&0&0&0&0&0&1.0674&0\\ 
		0&0&0&0&0&0&0&0&0&1.1251\\ 
		0&0&0&0&0&0&0&0&0&0\\ 
		0&0&0&0&0&0&0&0&0&0\\ 
		0&0&0&0&0&0&0&0&0&0\\ 
		0&0&0&0&0&0&0&0&0&0\\ 
		0&0&0&0&0&0&0&0&0&0\\ 
		0&0&0&0&0&0&0&0&0&0\\ 
		0&0&0&0&0&0&0&0&0&0\\ 
		0&0&0&0&0&0&0&0&0&0\\ 
		0&0&0&0&0&0&0&0&0&0
		\end{array}
    \right.\\\notag
	\tiny\left.
    \begin{array}{ccccccccc}
		0&0&0&0&0&0&0&0&0\\ 
		0&0&0&0&0&0&0&0&0\\ 
		0&0&0&0&0&0&0&0&0\\ 
		0&0&0&0&0&0&0&0&0\\ 
		0&0&0&0&0&0&0&0&0\\ 
		0&0&0&0&0&0&0&0&0\\ 
		0&0&0&0&0&0&0&0&0\\ 
		0&0&0&0&0&0&0&0&0\\ 
		0&0&0&0&0&0&0&0&0\\ 
		0&0&0&0&0&0&0&0&0\\ 
		1.1549&0&0&0&0&0&0&0&0\\ 
		0&1.2044&0&0&0&0&0&0&0\\ 
		0&0&1.2151&0&0&0&0&0&0\\ 
		0&0&0&1.2575&0&0&0&0&0\\ 
		0&0&0&0&1.2765&0&0&0&0\\ 
		0&0&0&0&0&1.3087&0&0&0\\ 
		0&0&0&0&0&0&1.4457&0&0\\ 
		0&0&0&0&0&0&0&1.5062&0\\ 
		0&0&0&0&0&0&0&0&1.7224
	\end{array}\right)
\end{align*}

\begin{align*}
  {\it{U}} = \tiny\left(
    \begin{array}{cccccccccc}
		0.017715&-0.10672&-0.044097&0.122&-0.025311&0.19466&0.017559&0.031527&0.026266&0.21037\\ 
		-0.026401&0.1139&0.2008&-0.022677&-0.010964&-0.14895&-0.026514&0.15558&-0.20015&-0.098011\\ 
		-0.043627&-0.052119&-0.00072014&-0.10949&-0.038862&0.17288&-0.39465&0.80446&-0.18569&-0.18224\\ 
		-0.030584&0.074443&0.011672&-0.26298&0.17022&0.1416&-0.098032&0.17843&0.75859&-0.0053179\\ 
		-0.039446&0.036163&0.068251&0.020913&0.068466&0.26719&-0.73391&-0.46851&-0.002116&-0.169\\ 
		0.089172&-0.098792&0.19551&0.29488&0.079416&0.47701&0.18966&-0.0071359&0.117&-0.51595\\ 
		-0.025317&-0.011293&0.038732&-0.031465&-0.085508&-0.36008&0.027458&-0.10708&0.17594&-0.60929\\ 
		0.0011027&-0.0198&0.14705&-0.21316&-0.91737&0.22007&0.038918&-0.081228&0.097578&0.033407\\ 
		0.76319&0.6302&-0.021246&0.05033&-0.045681&-0.00045645&-0.044612&0.043393&-0.024947&-0.017547\\ 
		0.00074598&-0.046756&-0.91121&0.19174&-0.17875&0.044271&-0.092648&0.0051454&0.011475&-0.13613\\ 
		-0.074069&0.014224&0.070437&0.0079671&-0.014823&0.050343&0.027964&-0.096178&-0.4691&-0.25038\\ 
		0.06539&0.011142&-0.20442&-0.7681&0.21815&0.35434&0.24868&-0.13419&-0.22622&-0.11049\\ 
		0.046025&-0.041143&-0.069197&0.13897&-0.0030214&0.02978&0.35991&0.086379&0.083122&-0.29167\\ 
		0.037146&-0.010418&-0.011386&0.072181&0.056061&0.21489&-0.05041&-0.065159&-0.0028081&0.010524\\ 
		-0.047937&0.049049&-0.055835&-0.32542&-0.080352&-0.44972&-0.15881&-0.029664&0.052968&-0.23783\\ 
		0.010796&-0.0039266&0.044461&-0.012649&-0.0056373&-0.059715&-0.098789&0.056668&-0.1002&-0.066757\\ 
		0.62245&-0.74028&0.045219&-0.085607&0.020765&-0.13957&-0.11092&-0.020046&0.0024831&0.024856\\ 
		0.0096252&-0.059944&0.044281&0.021085&-0.1288&0.10091&0.044254&0.11322&-0.030637&-0.032497\\ 
		0.018271&0.017284&-0.0087018&-0.044125&0.019124&-0.019187&0.047447&-0.0018637&0.049552&-0.064615
		\end{array}
    \right.\\\notag
	\tiny\left.
    \begin{array}{ccccccccc}
		0.082857&-0.15806&0.75862&0.0021224&0.14483&0.20443&-0.44765&0.082693&0.096616\\ 
		0.47082&-0.25243&0.29357&0.3813&-0.10399&-0.50924&0.22678&0.024794&-0.097322\\ 
		-0.19778&-0.054926&-0.010008&-0.13879&0.096864&0.052975&0.041919&0.074157&0.021214\\ 
		0.40435&-0.065337&-0.070703&0.031594&-0.20069&0.1992&0.011813&0.041946&-0.051519\\ 
		0.10007&0.030829&0.061232&-0.24457&0.11682&-0.2005&0.021505&0.019298&0.0077823\\ 
		-0.16937&-0.019614&-0.15506&0.35691&-0.065189&-0.15899&-0.3114&0.013433&0.01327\\ 
		-0.29298&-0.067685&0.42365&-0.14268&-0.22042&0.20279&0.23898&-0.032376&-0.085107\\ 
		0.013063&-0.12972&-0.064871&0.0039904&0.023215&-0.00010355&0.039433&-0.021546&0.0030931\\ 
		-0.017289&0.023866&0.026977&-0.038782&-0.011274&0.063822&-0.047156&0.030349&-0.0088835\\ 
		0.11484&-0.04961&-0.024382&0.15112&-0.14842&-0.087206&0.020425&-0.010955&-0.01788\\ 
		0.46516&-0.070846&-0.17587&-0.05184&-0.22602&0.55839&-0.088086&0.25772&0.040931\\ 
		-0.097806&-0.016255&0.13173&-0.02846&-0.017065&-0.10759&0.019908&-0.06522&-0.054306\\ 
		0.36497&-0.017927&-0.027561&-0.50398&0.56717&-0.14354&0.079827&-0.055484&-0.04685\\ 
		-0.0055343&0.10086&0.097097&0.47175&0.45639&0.40675&0.55067&-0.13322&-0.0070102\\ 
		0.049284&0.2067&-0.09872&0.34795&0.44218&0.015015&-0.44808&0.12669&0.059699\\ 
		0.15395&-0.038084&-0.029831&0.00081978&-0.082367&0.11937&-0.20776&-0.93369&0.068168\\ 
		0.099864&-0.04842&-0.052289&0.016401&-0.044216&-0.006795&0.052123&0.04154&-0.012598\\ 
		0.18662&0.90305&0.22004&-0.030171&-0.20461&-0.083954&0.040391&-0.010936&-0.006241\\ 
		0.023466&-0.020818&0.012148&-0.011965&-0.043652&-0.084373&0.14662&0.031707&0.9772
	\end{array}\right)
\end{align*}

Similarly, the covariance matrix for a 15-mW pulse entering into a PCF is given below, with a diagonal/unitary matrix obtained from eigenvalue decomposition.

\begin{align*}
  {\it{C}} = \tiny\left(
    \begin{array}{cccccccccc}
		3.1071&-0.018849&0.036555&-0.022881&0.052533&-0.09246&-0.060748&-0.18377&0.59718&0.23402\\ 
		-0.018849&2.9639&-0.0060527&0.034344&-0.032715&0.013188&-0.028785&0.32951&-0.083998&0.050002\\ 
		0.036555&-0.0060527&4.2768&0.0087372&-0.016274&-0.13086&0.10897&-0.038233&-0.027727&-0.19652\\ 
		-0.022881&0.034344&0.0087372&6.4697&0.0057684&0.074347&0.012015&-0.077824&-0.074677&0.092328\\ 
		0.052533&-0.032715&-0.016274&0.0057684&4.1836&0.062428&0.11021&0.035496&-0.053043&0.13403\\ 
		-0.09246&0.013188&-0.13086&0.074347&0.062428&3.6611&0.090594&0.26778&-0.09029&-0.14786\\ 
		-0.060748&-0.028785&0.10897&0.012015&0.11021&0.090594&5.1751&0.10201&0.11012&0.023962\\ 
		-0.18377&0.32951&-0.038233&-0.077824&0.035496&0.26778&0.10201&2.4282&0.089997&0.13224\\ 
		0.59718&-0.083998&-0.027727&-0.074677&-0.053043&-0.09029&0.11012&0.089997&3.3077&0.1572\\ 
		0.23402&0.050002&-0.19652&0.092328&0.13403&-0.14786&0.023962&0.13224&0.1572&2.0876\\ 
		-0.60657&-0.10001&0.03656&0.32376&-0.40464&0.1723&0.28964&0.21655&0.18525&0.16288\\ 
		0.016215&0.0763&0.12696&-0.56216&-0.07776&0.37141&0.049804&-0.11226&0.25524&0.22385\\ 
		0.051162&-0.28461&0.39978&-0.29639&0.59555&-0.43343&0.050902&0.34598&0.13194&-0.015651\\ 
		-0.52287&0.24481&-0.21822&0.4978&-0.077079&0.14359&-0.05023&-0.40404&0.21687&0.22257\\ 
		0.79616&-0.17462&-0.49923&0.44083&-0.69538&0.27109&-0.21335&0.49186&-0.30962&0.052642\\ 
		0.28506&-0.11199&0.59613&-0.30954&0.17235&-0.52207&0.45858&-0.47693&0.29825&-0.23046\\ 
		-0.75162&0.39798&0.40381&-0.29872&0.33771&-0.29226&-0.023192&0.35726&-0.62501&0.54294\\ 
		0.14035&-0.17059&0.12348&-0.69155&0.61453&-0.33489&0.078299&-0.76242&0.51234&-0.24223\\ 
		0.31256&0.029757&-1.0289&0.76672&-1.1191&0.52664&-0.7816&0.51083&-0.77751&-0.43546
		\end{array}
    \right.\\\notag
	\tiny\left.
    \begin{array}{ccccccccc}
		-0.60657&0.016215&0.051162&-0.52287&0.79616&0.28506&-0.75162&0.14035&0.31256\\ 
		-0.10001&0.0763&-0.28461&0.24481&-0.17462&-0.11199&0.39798&-0.17059&0.029757\\ 
		0.03656&0.12696&0.39978&-0.21822&-0.49923&0.59613&0.40381&0.12348&-1.0289\\ 
		0.32376&-0.56216&-0.29639&0.4978&0.44083&-0.30954&-0.29872&-0.69155&0.76672\\ 
		-0.40464&-0.07776&0.59555&-0.077079&-0.69538&0.17235&0.33771&0.61453&-1.1191\\ 
		0.1723&0.37141&-0.43343&0.14359&0.27109&-0.52207&-0.29226&-0.33489&0.52664\\ 
		0.28964&0.049804&0.050902&-0.05023&-0.21335&0.45858&-0.023192&0.078299&-0.7816\\ 
		0.21655&-0.11226&0.34598&-0.40404&0.49186&-0.47693&0.35726&-0.76242&0.51083\\ 
		0.18525&0.25524&0.13194&0.21687&-0.30962&0.29825&-0.62501&0.51234&-0.77751\\ 
		0.16288&0.22385&-0.015651&0.22257&0.052642&-0.23046&0.54294&-0.24223&-0.43546\\ 
		2.1268&0.26705&0.25566&-0.0082967&0.14595&0.077838&-0.55934&0.024708&0.017641\\ 
		0.26705&2.1861&0.259&0.26252&0.084848&0.12632&0.55689&-0.6268&-0.14008\\ 
		0.25566&0.259&2.2462&0.29776&0.26154&0.025204&-0.030746&0.21804&-0.29285\\ 
		-0.0082967&0.26252&0.29776&3.7221&0.32876&0.33478&0.057776&-0.12538&-0.45906\\ 
		0.14595&0.084848&0.26154&0.32876&2.8252&0.4448&0.38085&-0.022819&-0.65272\\ 
		0.077838&0.12632&0.025204&0.33478&0.4448&3.7847&0.42824&0.047635&-0.40341\\ 
		-0.55934&0.55689&-0.030746&0.057776&0.38085&0.42824&5.262&0.7174&-0.36252\\ 
		0.024708&-0.6268&0.21804&-0.12538&-0.022819&0.047635&0.7174&14.354&0.47308\\ 
		0.017641&-0.14008&-0.29285&-0.45906&-0.65272&-0.40341&-0.36252&0.47308&5.7507
	\end{array}\right)
\end{align*}

\begin{align*}
  {\it{V}} = \tiny\left(
    \begin{array}{cccccccccc}
		0.60807&0&0&0&0&0&0&0&0&0\\ 
		0&1.08&0&0&0&0&0&0&0&0\\ 
		0&0&1.2119&0&0&0&0&0&0&0\\ 
		0&0&0&1.8448&0&0&0&0&0&0\\ 
		0&0&0&0&2.1736&0&0&0&0&0\\ 
		0&0&0&0&0&2.6787&0&0&0&0\\ 
		0&0&0&0&0&0&2.8994&0&0&0\\ 
		0&0&0&0&0&0&0&3.157&0&0\\ 
		0&0&0&0&0&0&0&0&3.3397&0\\ 
		0&0&0&0&0&0&0&0&0&3.6583\\ 
		0&0&0&0&0&0&0&0&0&0\\ 
		0&0&0&0&0&0&0&0&0&0\\ 
		0&0&0&0&0&0&0&0&0&0\\ 
		0&0&0&0&0&0&0&0&0&0\\ 
		0&0&0&0&0&0&0&0&0&0\\ 
		0&0&0&0&0&0&0&0&0&0\\ 
		0&0&0&0&0&0&0&0&0&0\\ 
		0&0&0&0&0&0&0&0&0&0\\ 
		0&0&0&0&0&0&0&0&0&0
		\end{array}
    \right.\\\notag
	\tiny\left.
    \begin{array}{ccccccccc}
		0&0&0&0&0&0&0&0&0\\ 
		0&0&0&0&0&0&0&0&0\\ 
		0&0&0&0&0&0&0&0&0\\ 
		0&0&0&0&0&0&0&0&0\\ 
		0&0&0&0&0&0&0&0&0\\ 
		0&0&0&0&0&0&0&0&0\\ 
		0&0&0&0&0&0&0&0&0\\ 
		0&0&0&0&0&0&0&0&0\\ 
		0&0&0&0&0&0&0&0&0\\ 
		0&0&0&0&0&0&0&0&0\\ 
		3.8765&0&0&0&0&0&0&0&0\\ 
		0&4.1096&0&0&0&0&0&0&0\\ 
		0&0&4.5599&0&0&0&0&0&0\\ 
		0&0&0&4.6741&0&0&0&0&0\\ 
		0&0&0&0&5.0219&0&0&0&0\\ 
		0&0&0&0&0&5.9481&0&0&0\\ 
		0&0&0&0&0&0&6.3895&0&0\\ 
		0&0&0&0&0&0&0&8.0488&0\\ 
		0&0&0&0&0&0&0&0&14.638
	\end{array}\right)
\end{align*}

\begin{align*}
  {\it{U}} = \tiny\left(
    \begin{array}{cccccccccc}
		-0.40799&0.285&-0.069134&-0.02435&-0.11263&0.35808&-0.16669&0.17495&-0.15993&0.17428\\ 
		0.099599&-0.049009&-0.16029&0.1363&-0.23576&-0.37438&-0.58309&0.3872&-0.4739&-0.0086273\\ 
		0.2041&0.099172&0.085324&-0.13149&0.043347&0.18955&-0.16742&-0.01897&-0.064319&0.19071\\ 
		-0.092407&-0.11885&0.026921&0.10528&-0.043971&-0.036131&0.097774&0.010328&-0.039256&-0.039079\\ 
		0.22461&0.22561&0.13066&-0.040115&-0.060001&-0.076957&-0.071949&-0.11404&0.039366&0.40479\\ 
		-0.092204&-0.13935&-0.25274&-0.061762&0.066713&-0.086245&0.17739&0.21746&0.13405&0.6011\\ 
		0.086125&0.021048&0.0055473&0.028142&0.041231&0.17381&-0.081518&0.015214&-0.029029&-0.038808\\ 
		-0.33978&-0.11925&0.44074&-0.23993&0.34485&-0.28621&-0.27223&-0.40643&-0.16829&-0.034127\\ 
		0.26568&0.071469&-0.094364&0.094201&0.32723&-0.12222&0.39055&-0.13931&-0.66419&0.04969\\ 
		0.17698&-0.32757&-0.079147&-0.7248&-0.40049&0.21067&0.094964&-0.067067&-0.21259&-0.10694\\ 
		-0.16131&0.60817&-0.17923&-0.12653&-0.4454&-0.33731&0.082599&-0.3749&0.018634&-0.050639\\ 
		-0.031813&-0.18835&0.52679&0.4429&-0.52191&0.15684&0.1684&-0.14255&-0.17994&0.21593\\ 
		-0.079163&-0.35632&-0.54645&0.28333&-0.035149&0.15991&-0.27978&-0.59469&0.013867&0.078295\\ 
		-0.10718&0.19504&0.11289&-0.14943&0.20405&0.33952&-0.28949&-0.035477&-0.072867&0.26255\\ 
		0.57972&0.11942&0.085445&0.070042&0.027716&-0.086651&-0.21904&-0.10932&0.33286&-0.018973\\ 
		-0.12364&-0.23545&-0.0053139&-0.1579&-0.036628&-0.44345&0.093712&0.0118&0.12226&0.42004\\ 
		-0.10047&0.2058&-0.19747&0.079643&0.12227&0.093782&0.2428&-0.0017301&-0.13751&-0.017558\\ 
		-0.043567&-0.056297&0.061301&-0.0087414&-0.031096&-0.022065&-0.025881&0.0069306&0.014276&-0.018159\\ 
		0.28557&0.0972&-0.007489&-0.058623&0.0097338&0.13008&-0.045232&-0.20361&-0.16142&0.29702
		\end{array}
    \right.\\\notag
	\tiny\left.
    \begin{array}{ccccccccc}
		-0.28229&-0.52504&-0.083126&-0.21697&-0.15612&-0.23529&0.017947&0.04637&0.01212\\ 
		0.0018472&0.055904&0.063126&0.032898&0.047619&0.14925&0.010529&-0.001577&-0.015463\\ 
		-0.28499&0.42226&-0.67133&-0.10978&-0.078609&0.04114&-0.13923&-0.25982&0.014886\\ 
		-0.064614&-0.057639&-0.10504&0.053663&-0.065471&0.15448&-0.83841&0.43583&-0.084322\\ 
		0.14908&-0.3434&-0.015431&0.66727&-0.15086&0.024228&-0.13711&-0.23562&0.059738\\ 
		-0.4721&0.25238&0.26817&0.14773&0.17808&-0.018631&0.05397&0.13355&-0.036377\\ 
		0.045832&-0.12805&0.012147&0.0032451&0.86979&-0.2414&-0.25299&-0.22949&0.0067735\\ 
		-0.31039&-0.07792&0.037156&0.075301&0.12611&0.11097&0.075174&0.045952&-0.058705\\ 
		-0.073014&-0.025801&0.12021&-0.11763&-0.1411&-0.30997&-0.086314&-0.098494&0.039099\\ 
		-0.079911&-0.092845&0.15515&0.019984&-0.029419&0.086886&-0.048902&-0.065198&-0.018084\\ 
		-0.053877&0.23558&0.057614&-0.067907&0.10663&-0.093776&-0.074932&0.036982&-0.0072903\\ 
		-0.078197&0.090267&0.12327&-0.10142&0.037509&0.07276&0.078588&-0.10334&-0.043276\\ 
		-0.0080242&-0.0238&-0.011289&0.031795&-0.080239&-0.030677&-0.026062&-0.11207&0.020799\\ 
		0.3848&0.34473&0.47286&-0.18658&-0.11992&0.071417&-0.22679&-0.052008&-0.015799\\ 
		-0.37382&-0.24095&0.29713&-0.3841&-0.084578&0.077304&-0.10579&-0.0053325&-0.011575\\ 
		0.35264&-0.22138&-0.18946&-0.48706&0.0074899&-0.0082739&-0.10365&-0.22007&0.015448\\ 
		-0.069102&-0.15493&0.015721&-0.079496&0.14328&0.82094&0.019967&-0.25488&0.0715\\ 
		-0.049812&0.047866&0.038806&-0.013448&0.0051985&-0.020167&-0.059505&0.068355&0.98655\\ 
		0.22615&-0.099027&-0.21684&-0.089142&0.23204&0.14535&0.2937&0.67298&0.025738
	\end{array}\right)
\end{align*}

\end{widetext}

%%%%%%%%%% If using BibTeX:
\bibliographystyle{apsrev}
\bibliography{main}

\begin{thebibliography}{26}
\expandafter\ifx\csname natexlab\endcsname\relax\def\natexlab#1{#1}\fi
\expandafter\ifx\csname bibnamefont\endcsname\relax
  \def\bibnamefont#1{#1}\fi
\expandafter\ifx\csname bibfnamefont\endcsname\relax
  \def\bibfnamefont#1{#1}\fi
\expandafter\ifx\csname citenamefont\endcsname\relax
  \def\citenamefont#1{#1}\fi
\expandafter\ifx\csname url\endcsname\relax
  \def\url#1{\texttt{#1}}\fi
\expandafter\ifx\csname urlprefix\endcsname\relax\def\urlprefix{URL }\fi
\providecommand{\bibinfo}[2]{#2}
\providecommand{\eprint}[2][]{\url{#2}}

\bibitem[{\citenamefont{Dudley et~al.}(2006)\citenamefont{Dudley, Genty, and Coen}}]{PCF}
\bibinfo{author}{\bibfnamefont{J.~M.} \bibnamefont{Dudley}}, \bibinfo{author}{\bibfnamefont{G.}~\bibnamefont{Genty}}, \bibnamefont{and} \bibinfo{author}{\bibfnamefont{S.}~\bibnamefont{Coen}}, \bibinfo{journal}{Reviews of modern physics} \textbf{\bibinfo{volume}{78}}, \bibinfo{pages}{1135} (\bibinfo{year}{2006}).

\bibitem[{\citenamefont{Petersen et~al.}(2014)\citenamefont{Petersen, M{\o}ller, Kubat, Zhou, Dupont, Ramsay, Benson, Sujecki, Abdel-Moneim, Tang et~al.}}]{MIR}
\bibinfo{author}{\bibfnamefont{C.~R.} \bibnamefont{Petersen}}, \bibinfo{author}{\bibfnamefont{U.}~\bibnamefont{M{\o}ller}}, \bibinfo{author}{\bibfnamefont{I.}~\bibnamefont{Kubat}}, \bibinfo{author}{\bibfnamefont{B.}~\bibnamefont{Zhou}}, \bibinfo{author}{\bibfnamefont{S.}~\bibnamefont{Dupont}}, \bibinfo{author}{\bibfnamefont{J.}~\bibnamefont{Ramsay}}, \bibinfo{author}{\bibfnamefont{T.}~\bibnamefont{Benson}}, \bibinfo{author}{\bibfnamefont{S.}~\bibnamefont{Sujecki}}, \bibinfo{author}{\bibfnamefont{N.}~\bibnamefont{Abdel-Moneim}}, \bibinfo{author}{\bibfnamefont{Z.}~\bibnamefont{Tang}}, \bibnamefont{et~al.}, \bibinfo{journal}{Nature Photonics} \textbf{\bibinfo{volume}{8}}, \bibinfo{pages}{830} (\bibinfo{year}{2014}).

\bibitem[{\citenamefont{Humbert et~al.}(2006)\citenamefont{Humbert, Wadsworth, Leon-Saval, Knight, Birks, Russell, Lederer, Kopf, Wiesauer, Breuer et~al.}}]{OCT}
\bibinfo{author}{\bibfnamefont{G.}~\bibnamefont{Humbert}}, \bibinfo{author}{\bibfnamefont{W.}~\bibnamefont{Wadsworth}}, \bibinfo{author}{\bibfnamefont{S.}~\bibnamefont{Leon-Saval}}, \bibinfo{author}{\bibfnamefont{J.}~\bibnamefont{Knight}}, \bibinfo{author}{\bibfnamefont{T.}~\bibnamefont{Birks}}, \bibinfo{author}{\bibfnamefont{P.~S.~J.} \bibnamefont{Russell}}, \bibinfo{author}{\bibfnamefont{M.}~\bibnamefont{Lederer}}, \bibinfo{author}{\bibfnamefont{D.}~\bibnamefont{Kopf}}, \bibinfo{author}{\bibfnamefont{K.}~\bibnamefont{Wiesauer}}, \bibinfo{author}{\bibfnamefont{E.}~\bibnamefont{Breuer}}, \bibnamefont{et~al.}, \bibinfo{journal}{Optics express} \textbf{\bibinfo{volume}{14}}, \bibinfo{pages}{1596} (\bibinfo{year}{2006}).

\bibitem[{\citenamefont{Schenkel et~al.}(2005)\citenamefont{Schenkel, Paschotta, and Keller}}]{Compress}
\bibinfo{author}{\bibfnamefont{B.}~\bibnamefont{Schenkel}}, \bibinfo{author}{\bibfnamefont{R.}~\bibnamefont{Paschotta}}, \bibnamefont{and} \bibinfo{author}{\bibfnamefont{U.}~\bibnamefont{Keller}}, \bibinfo{journal}{JOSA B} \textbf{\bibinfo{volume}{22}}, \bibinfo{pages}{687} (\bibinfo{year}{2005}).

\bibitem[{\citenamefont{Jones et~al.}(2000)\citenamefont{Jones, Diddams, Ranka, Stentz, Windeler, Hall, and Cundiff}}]{OFC}
\bibinfo{author}{\bibfnamefont{D.~J.} \bibnamefont{Jones}}, \bibinfo{author}{\bibfnamefont{S.~A.} \bibnamefont{Diddams}}, \bibinfo{author}{\bibfnamefont{J.~K.} \bibnamefont{Ranka}}, \bibinfo{author}{\bibfnamefont{A.}~\bibnamefont{Stentz}}, \bibinfo{author}{\bibfnamefont{R.~S.} \bibnamefont{Windeler}}, \bibinfo{author}{\bibfnamefont{J.~L.} \bibnamefont{Hall}}, \bibnamefont{and} \bibinfo{author}{\bibfnamefont{S.~T.} \bibnamefont{Cundiff}}, \bibinfo{journal}{Science} \textbf{\bibinfo{volume}{288}}, \bibinfo{pages}{635} (\bibinfo{year}{2000}).

\bibitem[{\citenamefont{Ranka et~al.}(2000)\citenamefont{Ranka, Windeler, and Stentz}}]{SC_PCF}
\bibinfo{author}{\bibfnamefont{J.~K.} \bibnamefont{Ranka}}, \bibinfo{author}{\bibfnamefont{R.~S.} \bibnamefont{Windeler}}, \bibnamefont{and} \bibinfo{author}{\bibfnamefont{A.~J.} \bibnamefont{Stentz}}, \bibinfo{journal}{Optics letters} \textbf{\bibinfo{volume}{25}}, \bibinfo{pages}{25} (\bibinfo{year}{2000}).

\bibitem[{\citenamefont{Travers et~al.}(2011)\citenamefont{Travers, Chang, Nold, Joly, and Russell}}]{GAS}
\bibinfo{author}{\bibfnamefont{J.~C.} \bibnamefont{Travers}}, \bibinfo{author}{\bibfnamefont{W.}~\bibnamefont{Chang}}, \bibinfo{author}{\bibfnamefont{J.}~\bibnamefont{Nold}}, \bibinfo{author}{\bibfnamefont{N.~Y.} \bibnamefont{Joly}}, \bibnamefont{and} \bibinfo{author}{\bibfnamefont{P.~S.~J.} \bibnamefont{Russell}}, \bibinfo{journal}{JOSA B} \textbf{\bibinfo{volume}{28}}, \bibinfo{pages}{A11} (\bibinfo{year}{2011}).

\bibitem[{\citenamefont{Silva et~al.}(2012)\citenamefont{Silva, Austin, Thai, Baudisch, Hemmer, Faccio, Couairon, and Biegert}}]{filamentation}
\bibinfo{author}{\bibfnamefont{F.}~\bibnamefont{Silva}}, \bibinfo{author}{\bibfnamefont{D.}~\bibnamefont{Austin}}, \bibinfo{author}{\bibfnamefont{A.}~\bibnamefont{Thai}}, \bibinfo{author}{\bibfnamefont{M.}~\bibnamefont{Baudisch}}, \bibinfo{author}{\bibfnamefont{M.}~\bibnamefont{Hemmer}}, \bibinfo{author}{\bibfnamefont{D.}~\bibnamefont{Faccio}}, \bibinfo{author}{\bibfnamefont{A.}~\bibnamefont{Couairon}}, \bibnamefont{and} \bibinfo{author}{\bibfnamefont{J.}~\bibnamefont{Biegert}}, \bibinfo{journal}{Nature communications} \textbf{\bibinfo{volume}{3}}, \bibinfo{pages}{807} (\bibinfo{year}{2012}).

\bibitem[{\citenamefont{Lu et~al.}(2014)\citenamefont{Lu, Tsou, Chen, Chen, Cheng, Yang, Chen, Hsu, and Kung}}]{silica}
\bibinfo{author}{\bibfnamefont{C.-H.} \bibnamefont{Lu}}, \bibinfo{author}{\bibfnamefont{Y.-J.} \bibnamefont{Tsou}}, \bibinfo{author}{\bibfnamefont{H.-Y.} \bibnamefont{Chen}}, \bibinfo{author}{\bibfnamefont{B.-H.} \bibnamefont{Chen}}, \bibinfo{author}{\bibfnamefont{Y.-C.} \bibnamefont{Cheng}}, \bibinfo{author}{\bibfnamefont{S.-D.} \bibnamefont{Yang}}, \bibinfo{author}{\bibfnamefont{M.-C.} \bibnamefont{Chen}}, \bibinfo{author}{\bibfnamefont{C.-C.} \bibnamefont{Hsu}}, \bibnamefont{and} \bibinfo{author}{\bibfnamefont{A.~H.} \bibnamefont{Kung}}, \bibinfo{journal}{Optica} \textbf{\bibinfo{volume}{1}}, \bibinfo{pages}{400} (\bibinfo{year}{2014}).

\bibitem[{\citenamefont{Nakazawa et~al.}(1998)\citenamefont{Nakazawa, Tamura, Kubota, and Yoshida}}]{Coherence1}
\bibinfo{author}{\bibfnamefont{M.}~\bibnamefont{Nakazawa}}, \bibinfo{author}{\bibfnamefont{K.}~\bibnamefont{Tamura}}, \bibinfo{author}{\bibfnamefont{H.}~\bibnamefont{Kubota}}, \bibnamefont{and} \bibinfo{author}{\bibfnamefont{E.}~\bibnamefont{Yoshida}}, \bibinfo{journal}{Optical Fiber Technology} \textbf{\bibinfo{volume}{4}}, \bibinfo{pages}{215} (\bibinfo{year}{1998}).

\bibitem[{\citenamefont{Dudley and Coen}(2002)}]{Coherence2}
\bibinfo{author}{\bibfnamefont{J.~M.} \bibnamefont{Dudley}} \bibnamefont{and} \bibinfo{author}{\bibfnamefont{S.}~\bibnamefont{Coen}}, \bibinfo{journal}{IEEE Journal of selected topics in quantum electronics} \textbf{\bibinfo{volume}{8}}, \bibinfo{pages}{651} (\bibinfo{year}{2002}).

\bibitem[{\citenamefont{Zeylikovich et~al.}(2005)\citenamefont{Zeylikovich, Kartazaev, and Alfano}}]{Coherence3}
\bibinfo{author}{\bibfnamefont{I.}~\bibnamefont{Zeylikovich}}, \bibinfo{author}{\bibfnamefont{V.}~\bibnamefont{Kartazaev}}, \bibnamefont{and} \bibinfo{author}{\bibfnamefont{R.}~\bibnamefont{Alfano}}, \bibinfo{journal}{JOSA B} \textbf{\bibinfo{volume}{22}}, \bibinfo{pages}{1453} (\bibinfo{year}{2005}).

\bibitem[{\citenamefont{Corwin et~al.}(2003)\citenamefont{Corwin, Newbury, Dudley, Coen, Diddams, Weber, and Windeler}}]{Coherence4}
\bibinfo{author}{\bibfnamefont{K.~L.} \bibnamefont{Corwin}}, \bibinfo{author}{\bibfnamefont{N.~R.} \bibnamefont{Newbury}}, \bibinfo{author}{\bibfnamefont{J.~M.} \bibnamefont{Dudley}}, \bibinfo{author}{\bibfnamefont{S.}~\bibnamefont{Coen}}, \bibinfo{author}{\bibfnamefont{S.~A.} \bibnamefont{Diddams}}, \bibinfo{author}{\bibfnamefont{K.}~\bibnamefont{Weber}}, \bibnamefont{and} \bibinfo{author}{\bibfnamefont{R.}~\bibnamefont{Windeler}}, \bibinfo{journal}{Physical review letters} \textbf{\bibinfo{volume}{90}}, \bibinfo{pages}{113904} (\bibinfo{year}{2003}).

\bibitem[{\citenamefont{N{\"a}rhi et~al.}(2016)\citenamefont{N{\"a}rhi, Turunen, Friberg, and Genty}}]{Coherence5}
\bibinfo{author}{\bibfnamefont{M.}~\bibnamefont{N{\"a}rhi}}, \bibinfo{author}{\bibfnamefont{J.}~\bibnamefont{Turunen}}, \bibinfo{author}{\bibfnamefont{A.~T.} \bibnamefont{Friberg}}, \bibnamefont{and} \bibinfo{author}{\bibfnamefont{G.}~\bibnamefont{Genty}}, \bibinfo{journal}{Physical review letters} \textbf{\bibinfo{volume}{116}}, \bibinfo{pages}{243901} (\bibinfo{year}{2016}).

\bibitem[{\citenamefont{Halder et~al.}(2019)\citenamefont{Halder, Jukna, Koivurova, Dubietis, and Turunen}}]{Coherence6}
\bibinfo{author}{\bibfnamefont{A.}~\bibnamefont{Halder}}, \bibinfo{author}{\bibfnamefont{V.}~\bibnamefont{Jukna}}, \bibinfo{author}{\bibfnamefont{M.}~\bibnamefont{Koivurova}}, \bibinfo{author}{\bibfnamefont{A.}~\bibnamefont{Dubietis}}, \bibnamefont{and} \bibinfo{author}{\bibfnamefont{J.}~\bibnamefont{Turunen}}, \bibinfo{journal}{Photonics Research} \textbf{\bibinfo{volume}{7}}, \bibinfo{pages}{1345} (\bibinfo{year}{2019}).

\bibitem[{\citenamefont{Sulzer et~al.}(2020)\citenamefont{Sulzer, Beckh, Liehl, Huster, Keller, Cimander, Henzler, Traum, Riek, Seletskiy et~al.}}]{Coherence7}
\bibinfo{author}{\bibfnamefont{P.}~\bibnamefont{Sulzer}}, \bibinfo{author}{\bibfnamefont{C.}~\bibnamefont{Beckh}}, \bibinfo{author}{\bibfnamefont{A.}~\bibnamefont{Liehl}}, \bibinfo{author}{\bibfnamefont{J.}~\bibnamefont{Huster}}, \bibinfo{author}{\bibfnamefont{K.~R.} \bibnamefont{Keller}}, \bibinfo{author}{\bibfnamefont{M.}~\bibnamefont{Cimander}}, \bibinfo{author}{\bibfnamefont{P.}~\bibnamefont{Henzler}}, \bibinfo{author}{\bibfnamefont{C.}~\bibnamefont{Traum}}, \bibinfo{author}{\bibfnamefont{C.}~\bibnamefont{Riek}}, \bibinfo{author}{\bibfnamefont{D.~V.} \bibnamefont{Seletskiy}}, \bibnamefont{et~al.}, \bibinfo{journal}{Optics Letters} \textbf{\bibinfo{volume}{45}}, \bibinfo{pages}{4714} (\bibinfo{year}{2020}).

\bibitem[{\citenamefont{Friberg et~al.}(1996)\citenamefont{Friberg, Machida, Werner, Levanon, and Mukai}}]{soliton_filter}
\bibinfo{author}{\bibfnamefont{S.}~\bibnamefont{Friberg}}, \bibinfo{author}{\bibfnamefont{S.}~\bibnamefont{Machida}}, \bibinfo{author}{\bibfnamefont{M.}~\bibnamefont{Werner}}, \bibinfo{author}{\bibfnamefont{A.}~\bibnamefont{Levanon}}, \bibnamefont{and} \bibinfo{author}{\bibfnamefont{T.}~\bibnamefont{Mukai}}, \bibinfo{journal}{Physical review letters} \textbf{\bibinfo{volume}{77}}, \bibinfo{pages}{3775} (\bibinfo{year}{1996}).

\bibitem[{\citenamefont{Hirosawa et~al.}(2005)\citenamefont{Hirosawa, Furumochi, Tada, Kannari, Takeoka, and Sasaki}}]{Hirosawa}
\bibinfo{author}{\bibfnamefont{K.}~\bibnamefont{Hirosawa}}, \bibinfo{author}{\bibfnamefont{H.}~\bibnamefont{Furumochi}}, \bibinfo{author}{\bibfnamefont{A.}~\bibnamefont{Tada}}, \bibinfo{author}{\bibfnamefont{F.}~\bibnamefont{Kannari}}, \bibinfo{author}{\bibfnamefont{M.}~\bibnamefont{Takeoka}}, \bibnamefont{and} \bibinfo{author}{\bibfnamefont{M.}~\bibnamefont{Sasaki}}, \bibinfo{journal}{Physical review letters} \textbf{\bibinfo{volume}{94}}, \bibinfo{pages}{203601} (\bibinfo{year}{2005}).

\bibitem[{\citenamefont{Sp{\"a}lter et~al.}(1998)\citenamefont{Sp{\"a}lter, Korolkova, K{\"o}nig, Sizmann, and Leuchs}}]{Soliton}
\bibinfo{author}{\bibfnamefont{S.}~\bibnamefont{Sp{\"a}lter}}, \bibinfo{author}{\bibfnamefont{N.}~\bibnamefont{Korolkova}}, \bibinfo{author}{\bibfnamefont{F.}~\bibnamefont{K{\"o}nig}}, \bibinfo{author}{\bibfnamefont{A.}~\bibnamefont{Sizmann}}, \bibnamefont{and} \bibinfo{author}{\bibfnamefont{G.}~\bibnamefont{Leuchs}}, \bibinfo{journal}{Physical review letters} \textbf{\bibinfo{volume}{81}}, \bibinfo{pages}{786} (\bibinfo{year}{1998}).

\bibitem[{\citenamefont{Opatrn{\`y} et~al.}(2002)\citenamefont{Opatrn{\`y}, Korolkova, and Leuchs}}]{ModeStructure}
\bibinfo{author}{\bibfnamefont{T.}~\bibnamefont{Opatrn{\`y}}}, \bibinfo{author}{\bibfnamefont{N.}~\bibnamefont{Korolkova}}, \bibnamefont{and} \bibinfo{author}{\bibfnamefont{G.}~\bibnamefont{Leuchs}}, \bibinfo{journal}{Physical Review A} \textbf{\bibinfo{volume}{66}}, \bibinfo{pages}{053813} (\bibinfo{year}{2002}).

\bibitem[{\citenamefont{Roslund et~al.}(2014)\citenamefont{Roslund, De~Araujo, Jiang, Fabre, and Treps}}]{WDM}
\bibinfo{author}{\bibfnamefont{J.}~\bibnamefont{Roslund}}, \bibinfo{author}{\bibfnamefont{R.~M.} \bibnamefont{De~Araujo}}, \bibinfo{author}{\bibfnamefont{S.}~\bibnamefont{Jiang}}, \bibinfo{author}{\bibfnamefont{C.}~\bibnamefont{Fabre}}, \bibnamefont{and} \bibinfo{author}{\bibfnamefont{N.}~\bibnamefont{Treps}}, \bibinfo{journal}{Nature Photonics} \textbf{\bibinfo{volume}{8}}, \bibinfo{pages}{109} (\bibinfo{year}{2014}).

\bibitem[{\citenamefont{Wakui et~al.}(2014)\citenamefont{Wakui, Eto, Benichi, Izumi, Yanagida, Ema, Numata, Fukuda, Takeoka, and Sasaki}}]{Wakui-san}
\bibinfo{author}{\bibfnamefont{K.}~\bibnamefont{Wakui}}, \bibinfo{author}{\bibfnamefont{Y.}~\bibnamefont{Eto}}, \bibinfo{author}{\bibfnamefont{H.}~\bibnamefont{Benichi}}, \bibinfo{author}{\bibfnamefont{S.}~\bibnamefont{Izumi}}, \bibinfo{author}{\bibfnamefont{T.}~\bibnamefont{Yanagida}}, \bibinfo{author}{\bibfnamefont{K.}~\bibnamefont{Ema}}, \bibinfo{author}{\bibfnamefont{T.}~\bibnamefont{Numata}}, \bibinfo{author}{\bibfnamefont{D.}~\bibnamefont{Fukuda}}, \bibinfo{author}{\bibfnamefont{M.}~\bibnamefont{Takeoka}}, \bibnamefont{and} \bibinfo{author}{\bibfnamefont{M.}~\bibnamefont{Sasaki}}, \bibinfo{journal}{Scientific reports} \textbf{\bibinfo{volume}{4}}, \bibinfo{pages}{1} (\bibinfo{year}{2014}).

\bibitem[{\citenamefont{Ng et~al.}(2023)\citenamefont{Ng, Yanagimoto, Jankowski, Fejer, and Mabuchi}}]{GSSF}
\bibinfo{author}{\bibfnamefont{E.}~\bibnamefont{Ng}}, \bibinfo{author}{\bibfnamefont{R.}~\bibnamefont{Yanagimoto}}, \bibinfo{author}{\bibfnamefont{M.}~\bibnamefont{Jankowski}}, \bibinfo{author}{\bibfnamefont{M.}~\bibnamefont{Fejer}}, \bibnamefont{and} \bibinfo{author}{\bibfnamefont{H.}~\bibnamefont{Mabuchi}}, \bibinfo{journal}{arXiv preprint arXiv:2307.05464}  (\bibinfo{year}{2023}).

\bibitem[{\citenamefont{Weedbrook et~al.}(2012)\citenamefont{Weedbrook, Pirandola, Garc{\'\i}a-Patr{\'o}n, Cerf, Ralph, Shapiro, and Lloyd}}]{GQI}
\bibinfo{author}{\bibfnamefont{C.}~\bibnamefont{Weedbrook}}, \bibinfo{author}{\bibfnamefont{S.}~\bibnamefont{Pirandola}}, \bibinfo{author}{\bibfnamefont{R.}~\bibnamefont{Garc{\'\i}a-Patr{\'o}n}}, \bibinfo{author}{\bibfnamefont{N.~J.} \bibnamefont{Cerf}}, \bibinfo{author}{\bibfnamefont{T.~C.} \bibnamefont{Ralph}}, \bibinfo{author}{\bibfnamefont{J.~H.} \bibnamefont{Shapiro}}, \bibnamefont{and} \bibinfo{author}{\bibfnamefont{S.}~\bibnamefont{Lloyd}}, \bibinfo{journal}{Reviews of Modern Physics} \textbf{\bibinfo{volume}{84}}, \bibinfo{pages}{621} (\bibinfo{year}{2012}).

\bibitem[{\citenamefont{Hosaka et~al.}(2016)\citenamefont{Hosaka, Kawamori, and Kannari}}]{Multimode}
\bibinfo{author}{\bibfnamefont{A.}~\bibnamefont{Hosaka}}, \bibinfo{author}{\bibfnamefont{T.}~\bibnamefont{Kawamori}}, \bibnamefont{and} \bibinfo{author}{\bibfnamefont{F.}~\bibnamefont{Kannari}}, \bibinfo{journal}{Physical Review A} \textbf{\bibinfo{volume}{94}}, \bibinfo{pages}{053833} (\bibinfo{year}{2016}).

\bibitem[{\citenamefont{Ferrini et~al.}(2013)\citenamefont{Ferrini, Gazeau, Coudreau, Fabre, and Treps}}]{GQC}
\bibinfo{author}{\bibfnamefont{G.}~\bibnamefont{Ferrini}}, \bibinfo{author}{\bibfnamefont{J.-P.} \bibnamefont{Gazeau}}, \bibinfo{author}{\bibfnamefont{T.}~\bibnamefont{Coudreau}}, \bibinfo{author}{\bibfnamefont{C.}~\bibnamefont{Fabre}}, \bibnamefont{and} \bibinfo{author}{\bibfnamefont{N.}~\bibnamefont{Treps}}, \bibinfo{journal}{New Journal of Physics} \textbf{\bibinfo{volume}{15}}, \bibinfo{pages}{093015} (\bibinfo{year}{2013}).

\end{thebibliography}

%\section{Auther contribution}

%A.H, S.N., M.T, A.O, M.T., and F.K. conceived the project. A.H. designed and performed the experiments. In addition, A.H. analyzed the experimental results. S.N., A.O., and M.T assisted construction of experimental setup. S.N. made electric circuits for the homodyne detection system. A.O. and M.T. assisted the data acquisition. A.H, S.N., M.T and F.K. wrote the manuscript. M.T and F.K. supervised the research team and interpreted the results. All authors discussed the results.

\end{document}